\documentclass[twocolumn,preprintnumbers,superscriptaddress,amsmath,amssymb]{revtex4}
\usepackage{graphicx}% Include figure files
\usepackage{dcolumn}% Align table columns on decimal point
\usepackage{bm}% bold math
\usepackage{amssymb}
\usepackage{amssymb}
\usepackage{verbatim}
\usepackage{color}
\usepackage[colorlinks=true,linkcolor=blue,citecolor=blue,urlcolor=blue]{hyperref}
\def\>{\rangle}

\begin{document}

\title{A solid-state quantum memory based on a continuous optoacoustic system}% Force line breaks with\\
\author{Changlong Zhu}
\affiliation{Max Planck Institute for the Science of Light, Staudtstra{\ss}e 2,
D-91058 Erlangen, Germany}
\affiliation{Institute of Photonics, Leibniz University Hannover, Welfengarten 1A, 30167 Hannover, Germany}
\author{Claudiu Genes}
\affiliation{TU Darmstadt, Institute for Applied Physics, Hochschulstraße 4A, D-64289 Darmstadt, Germany}
\author{Birgit Stiller}\email{birgit.stiller@mpl.mpg.de}
\affiliation{Max Planck Institute for the Science of Light, Staudtstra{\ss}e 2,
D-91058 Erlangen, Germany}
\affiliation{Institute of Photonics, Leibniz University Hannover, Welfengarten 1A, 30167 Hannover, Germany}
\affiliation{Department of Physics, Friedrich-Alexander-Universit\"{a}t Erlangen-N{\"u}rnberg, Staudtstra{\ss}e 7,
D-91058 Erlangen, Germany}
\date{\today}

\begin{abstract}
Quantum memories for optical states are essential resources for quantum communication and information processing. We propose a quantum memory protocol based on coherent photon–phonon transduction in a Brillouin-active optical waveguide supporting traveling acoustic modes. A pulsed pump drives an effective beam-splitter interaction between optical and acoustic fields, enabling the mapping of a propagating optical quantum state onto a traveling phononic excitation and its subsequent retrieval on demand. Using a continuum optoacoustic model, we show that the protocol enables broadband quantum state storage in a distributed medium without relying on discrete cavity modes. Analytical and numerical results demonstrate high-fidelity storage and retrieval of squeezed and entangled states under experimentally realistic parameters. The memory bandwidth is set by the Brillouin interaction and can reach hundreds of MHz. Our results identify continuum Brillouin optomechanical systems as a scalable platform for broadband quantum memories and multimode quantum signal processing.

\end{abstract}

\pacs{42.50.Ar, 42.50.Lc, 42.72.-g}
%\pacs{05.45.Xt,07.10.Cm,45.50.Wk,42.65.-k}

\maketitle
Quantum memories~\cite{lvovsky2009,afzelius2015} enable the reversible storage of optical quantum states and are essential for quantum networks~\cite{kimble2008,wehner2018}, quantum imaging~\cite{mazelanik2021}, quantum metrology~\cite{zaiser2016,degen2017}, and space-based tests of fundamental physics~\cite{mol2023}. Their performance is characterized by fidelity, storage time, retrieval efficiency, bandwidth, and multimode capacity. A wide range of physical platforms has been explored~\cite{heshami2016,lei2023}, including atomic ensembles~\cite{julsgaard2004}, single atoms~\cite{li2016}, rare-earth–ion-doped solids~\cite{clausen2012,gundougan2012,zhou2012}, nitrogen-vacancy centers in diamond~\cite{shandilya2021}, and optomechanical systems~\cite{wallucks2020,liu2023coherent}. Despite substantial progress, achieving a quantum memory that simultaneously combines large bandwidth, high fidelity, and scalability remains challenging. Cavity optomechanical platforms~\cite{aspelmeyer2014,metcalfe2014,barzanjeh2022} have demonstrated promising results, including a Duan–Lukin–Cirac–Zoller-type quantum memory with a lifetime of $T_1 \approx 2\,\mathrm{ms}$ at $15\,\mathrm{mK}$~\cite{wallucks2020}. However, their reliance on discrete cavity modes and deep cryogenic operation limits spectral bandwidth and integration with scalable photonic architectures.

%%%%%%%%%%%%%%%%%%%%%%%%%%%%%%%
%%%%%%%%%%%%%%%%%%%%%%%%%%%%%%%
\begin{figure}[b]
	\centerline{
		\includegraphics[clip, width=0.95\linewidth]{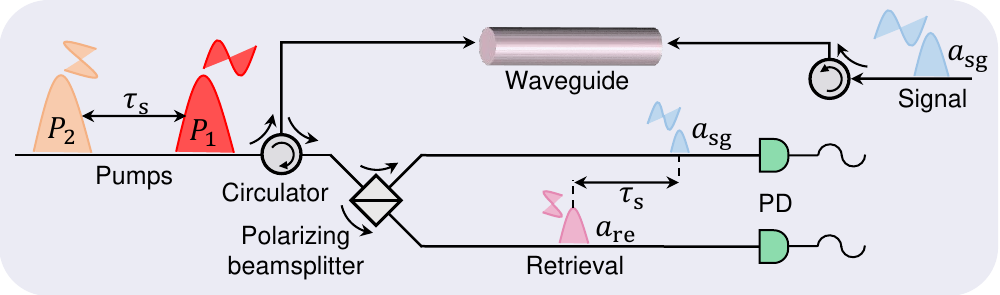}}
	\caption{Schematic of the quantum memory protocol in a Brillouin-active waveguide. 
An input signal field $a_{\rm sg}$ enters the waveguide and counter-propagates with a first pump pulse $P_1$, which drives the backward Brillouin anti-Stokes interaction and coherently maps the optical state onto a propagating acoustic excitation. After a storage time $\tau_{\rm s}$, a second pump pulse $P_2$ converts the phononic excitation back into an optical retrieval field $a_{\rm re}$. The two pump pulses are orthogonally polarized, allowing the retrieved signal to be separated from the transmitted light using a polarizing beamsplitter and detected with photodetectors (PD). Circulators direct the optical fields into and out of the Brillouin-active waveguide.
	}\label{Fig1}
\end{figure}
%%%%%%%%%%%%%%%%%%%%%%%%%%%%%%%
%%%%%%%%%%%%%%%%%%%%%%%%%%%%%%%
A promising route toward overcoming these limitations is offered by continuum optomechanical systems based on Brillouin scattering in optical waveguides~\cite{van2016,rakich2018}. In such systems, optical photons interact with propagating acoustic phonons that are continuously accessible along the waveguide~\cite{otterstrom2018,zhu2024}. This traveling-wave architecture naturally supports large optical and acoustic bandwidths and enables parallel signal processing across multiple frequency channels~\cite{stiller2019,becker2024}. Rapid progress in nanofabrication has led to chip-scale waveguides with large Brillouin gain~\cite{eggleton2019,gyger2020,botter2022,rodrigues2025}, enabling a variety of quantum optomechanical experiments~\cite{zoubi2017,otterstrom2018,zhang2023,zhu2023,johnson2023,blazquez2024,zhu2024,zoubi2024,cryer2025}. Recent demonstrations include strong photon–phonon coupling in nonlinear fibers at cryogenic temperatures~\cite{martinez2025}, ground-state operation of GHz acoustic modes at sub-kelvin temperatures~\cite{cryer2025}, and protocols for generating optical–phononic entanglement~\cite{zhu2024}. These developments establish Brillouin-active waveguides as a promising platform for broadband quantum photonic technologies. However, a complete protocol for storing and retrieving quantum optical states in such continuum optomechanical systems has not yet been demonstrated.

In this work, we propose a quantum memory protocol for optical states in Brillouin-active waveguides (Fig.~\ref{Fig1}). The protocol relies on coherent photon–phonon transduction mediated by an effective beam-splitter interaction between optical photons and traveling acoustic phonons. A first pump pulse $P_1$ of duration $\tau_1$ drives the backward Brillouin anti-Stokes process, mapping the quantum state of an input signal field $a_{\rm sg}$ onto a propagating phononic excitation. After a programmable storage time $\tau_{\rm s}$, a second pump pulse $P_2$ with duration $\tau_2$ converts the phononic state back into an optical retrieval field $a_{\rm re}$. The two pump pulses are arranged with orthogonal polarizations, allowing the retrieved field to be separated from the transmitted light using a polarizing beamsplitter. Using experimentally realistic parameters for state-of-the-art Brillouin platforms, we show that nonclassical optical states, including squeezed and entangled states, can be stored and retrieved with high fidelity at sub-kelvin temperatures while remaining robust at elevated temperatures.\\

\noindent \textbf{Quantum transduction via the Brillouin anti-Stokes process}.—
In optical waveguides, backward Brillouin anti-Stokes scattering converts a pump photon with frequency $\omega_{\rm p}$ into a higher-frequency anti-Stokes photon $\omega_{\rm as}=\omega_{\rm p}+\Omega_{\rm ac}$ through the absorption of an acoustic phonon of frequency $\Omega_{\rm ac}$. 
In the undepleted pump regime, this three-wave interaction can be linearized and described by an effective optoacoustic beam-splitter interaction between anti-Stokes photons and acoustic phonons with coupling strength $g$~\cite{zhu2024} (see Fig.~\ref{Fig1}). 
This interaction enables coherent exchange of quantum states between optical and acoustic modes and thus provides the mechanism for photon–phonon quantum transduction.

We briefly summarize the dynamics of the linearized Brillouin anti-Stokes process driven by a cw pump. 
In a traveling-wave waveguide without a cavity, the optical anti-Stokes and acoustic fields form continua of photon and phonon modes~\cite{kharel2016,sipe2016,zoubi2016} with envelope operators
\begin{subequations}
\begin{align}
a_{\rm as}(z,t) &= \frac{1}{\sqrt{2\pi}}\!\int dk\, a(k,t)e^{-ikz}, \\
b_{\rm ac}(z,t) &= \frac{1}{\sqrt{2\pi}}\!\int dk\, b(k,t)e^{-ikz}.
\end{align}
\end{subequations}
Moving to momentum space, the linearized dynamics read
\begin{subequations}
\begin{align}
\dot a &= \left(-\frac{\gamma}{2}+i\Delta_{\rm as}\right)a
         -ig\, b + \sqrt{\gamma}\,\xi_{\rm as}, \\
\dot b &= \left(-\frac{\Gamma}{2}+i\Delta_{\rm ac}\right)b
         -ig\, a + \sqrt{\Gamma}\,\xi_{\rm ac}.
\end{align}
\end{subequations}
Here $\gamma$ and $\Gamma$ denote the optical and acoustic dissipation rates. 
The detunings $\Delta_{\rm as}=kv_{\rm o}$ and $\Delta_{\rm ac}=kv_{\rm ac}$ arise from wave-number–dependent frequency shifts of the anti-Stokes and acoustic modes with group velocities $v_{\rm o}$ and $v_{\rm ac}$, respectively. 
The coupling constant $g$ is taken to be real and positive without loss of generality. 
The quantum noise operators $\xi_{\rm as}$ and $\xi_{\rm ac}$ satisfy
\begin{subequations}
\begin{align}
\langle \xi_{\rm as}(t_1)\xi_{\rm as}^\dagger(t_2)\rangle
&= \delta(t_1-t_2), \\
\langle \xi_{\rm ac}(t_1)\xi_{\rm ac}^\dagger(t_2)\rangle
&= (n_{\rm th}+1)\,\delta(t_1-t_2).
\end{align}
\end{subequations}
where
\begin{equation}
n_{\rm th}=\left(e^{\hbar\Omega_{\rm ac}/k_{\rm B}T_{\rm en}}-1\right)^{-1}.
\end{equation}
The thermal occupation of optical modes is negligible due to the high optical frequency ($\sim193\,{\rm THz}$).
%%%%%%%%%%%%%%%%%%%%%%%%%%%%%%%
%%%%%%%%%%%%%%%%%%%%%%%%%%%%%%%

\begin{figure}[t]
	\centerline{
		\includegraphics[clip, width=\linewidth]{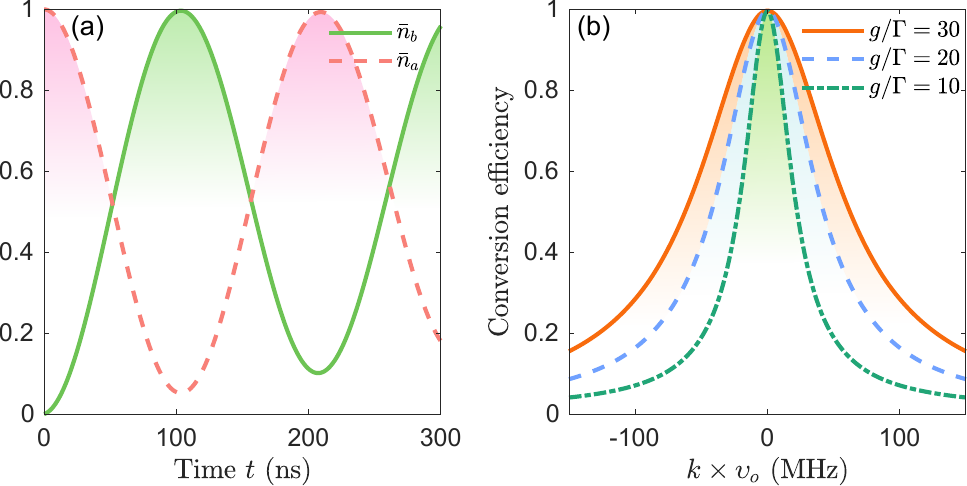}}
	\caption{(a) Time evolution of the acoustic phonon occupation number $\bar{n}_b(t)=n_b/n_{a,0}$
		and anti-Stokes photon occupation number $\bar{n}_{a}(t)=n_{a}/n_{a,0}$ in the strong coupling
		regime ($g/\Gamma=15$). (b) Conversion efficiency of the continuum optoacoustic transduction versus the 
		wave number $k$ in the strong coupling regime.  
	}\label{Fig2}
\end{figure}
%%%%%%%%%%%%%%%%%%%%%%%%%%%%%%%
%%%%%%%%%%%%%%%%%%%%%%%%%%%%%%%

The coupled equations for the envelope operators can be readily analytically solved. In the strong-coupling regime ($g>\gamma,\Gamma$), the average phonon number admits the approximate analytical form
\begin{equation}\label{Transferred phonon number}
\begin{aligned}
n_b(t) \approx &\;
\frac{n_{a,0}}{2} e^{-\frac{\gamma+\Gamma}{2}t}\!\left[1-\cos(2gt)\right]  \\
&+ \frac{n_{b,0}}{2} e^{-\frac{\gamma+\Gamma}{2}t}\!\left[1+\cos(2gt)\right]  \\
&+ \left(1-e^{-\frac{\gamma+\Gamma}{2}t}\right)n_{\rm th}.
\end{aligned}
\end{equation}
Here $n_{a,0}$ and $n_{b,0}$ denote the initial photon and phonon populations of the anti-Stokes and acoustic modes, respectively. 
We assume that the wave-number–induced frequency shifts lie within the acoustic linewidth, $\Delta_{\rm as,ac}\ll\Gamma$, and that 
$\Delta_{\rm ac}\ll\Delta_{\rm as}$, consistent with the large difference between acoustic and optical propagation velocities 
($\upsilon_{\rm ac}\ll\upsilon_{\rm as}$). 

Equation~(\ref{Transferred phonon number}) shows that energy is coherently exchanged between photons and phonons via 
Rabi oscillations at frequency $\sim 2g$. This behavior is confirmed by numerical simulations in Fig.~\ref{Fig2}(a), where we used the following experimentally realistic parameters~\cite{eggleton2019,morrison2017}: 
$\Gamma/2\pi=1\,\mathrm{MHz}$, $\gamma/2\pi=0.2\,\mathrm{MHz}$, 
$\Omega_{\rm ac}/2\pi=7.6\,\mathrm{GHz}$, environmental temperature $T_{\rm en}=4\,\mathrm{K}$, 
and detuning $\Delta_{\rm as}=0.2\Gamma$.
At the optimal time, $t_{\rm opt}\sim \pi/2g$, the swap is complete and its efficiency is only limited because of the decoherence brought on by the last term in Eq.~(\ref{Transferred phonon number}). In the strong coupling regime, the photon-phonon conversion efficiency~\cite{lauk2020,wang2022}, quantifying the fraction of anti-Stokes photon number that 
can be converted into acoustic phonons, reaches its maximum at $t_{\rm opt}$ and can be approximately 
expressed as 
\begin{equation}
    \eta_\text{swap}=1-\frac{\pi(\gamma+\Gamma)}{4g}.
\end{equation}
It indicates that a high photon-phonon 
conversion efficiency is achievable in the strong coupling regime for continuously
accessible groups of photons and phonons, enabling quantum transduction over a broad bandwidth. This is illustrated in
Fig.~\ref{Fig2}(b), which shows the conversion efficiency over a large interval of wave number $k$ at an environment temperature of $4~{\rm K}$.
This capability opens new opportunities for quantum technologies, including quantum information processing~\cite{kimble2008}, 
quantum computing~\cite{andersen2015}, and quantum sensing~\cite{ye2024}.

%%%%%%%%%%%%%%%%%%%%%%%%%%%%%%%
%%%%%%%%%%%%%%%%%%%%%%%%%%%%%%%
\begin{figure}[t]
	\centerline{
		\includegraphics[clip, width=\linewidth]{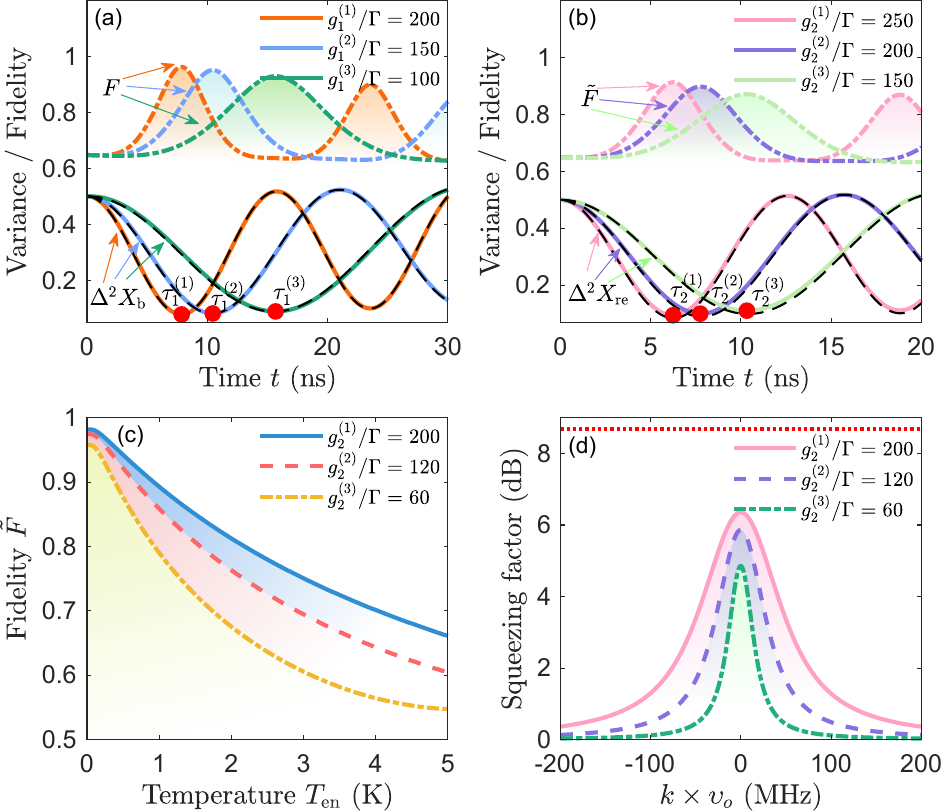}}
	\caption{ (a) Time evolution of quadrature variance $\Delta^2 X_{\rm b}$ and fidelity 
		$F[\mathcal{V}_{sg}(0),\mathcal{V}_{\rm b}(t)]$ at $T_{\rm en}=1~{\rm K}$ in the 
		writing process for various coupling ratio $g_1^{(i)}/\Gamma$ ($i=1,2,3$), where red points denote the optimal
		time for maximum quantum transduction and black dashed lines correspond to 
		approximate analytical solutions. (b) Time evolution of $\Delta^2 X_{\rm re}$
		and $\tilde{F}[\mathcal{V}_{sg}(0),\mathcal{V}_{\rm re}(t)]$ in the readout process
		for various ratios $g_2^{(i)}/\Gamma$ with a storage time of $\tau_{\rm s}=5~{\rm ns}$.
		(c) Variation of $\tilde{F}$ versus $T_{\rm en}$ for various ratios $g_2/\Gamma$ ($g_1=g_2$)
		after a storage time of $\tau_{\rm s}=5~{\rm ns}$. (d) Continuum memory versus 
		the wave number $k$ in the strong coupling regime ($g_1=g_2$) at $T_{\rm en}=1~{\rm K}$, where red 
		dotted line denotes the initial squeezing factor of the signal light and $\tau_{\rm s}=10~{\rm ns}$.  
	}\label{Fig3}
\end{figure}
%%%%%%%%%%%%%%%%%%%%%%%%%%%%%%%
%%%%%%%%%%%%%%%%%%%%%%%%%%%%%%%

\noindent \textbf{Quantum storage of squeezed states}.---We consider quantum memory of squeezed states implemented via quantum transduction in 
Brillouin-active waveguides. We assume that a signal light $a_{\rm sg}$ is initially prepared in a squeezed state. Since optical and acoustic states can be
interconverted through the quantum transduction, the squeezed state is first mapped from the signal light $a_{\rm sg}$ onto acoustic phonons $b$
by a pulsed pump of duration $\tau_1$ in the writing process. After being stored in the acoustic phonons for a finite duration $\tau_{\text s}$, the state is subsequently retrieved by mapping it back onto a retrieval light $a_{\text re}$ by a second pulsed pump of duration $\tau_2$ in the readout process, as illustrated in Figs.~\ref{Fig1}(b) and (c). Here, $a_{\rm sg}$,
$a_{\rm re}$, and $b$ correspond to annihilation operators of photons and phonons with the wave number $k$, respectively. For simplification,
we consider a squeezed vacuum state $S(r)\left |0 \right\rangle $ for the signal photons, ground state for acoustic phonons, and vacuum state for
retrieval photons at initial time, where the unitary
phase-free squeezing operator $S(r)$ with a squeezing degree $r$ can be expressed as $S(r)=\exp\left[\frac{r}{2}( a_{\rm sg}^2 - (a_{\rm sg}^{\dagger})^2 )\right]$.
The squeezed vacuum state belongs to the family of Gaussian states and can be completely characterized by its covariance matrix 
$\mathcal{V}_{\rm sg} (0)=1/2{\rm diag}\left\{\exp(-2r), \exp(2r)\right\}$~\cite{adesso2008,weedbrook2012}. During the writing process, the variance of the squeezed 
acoustic quadrature via the quantum transduction can be given by a simple and approximate expression
\begin{equation} 
\Delta^2 X_{\rm b} \approx  \frac{1}{2} - \frac{1 - e^{-2r}}{2} e^{-\frac{\Gamma}{2} t} \sin^2(g_1t) + \frac{\Gamma n_{\rm th}}{2}t,
\end{equation}
where $g_1$ denotes the effective optoacoustic coupling strength. The Rabi oscillation of $\Delta^2 X_{\rm b}$ implies that the minimum value can be obtained at the end of the first half
Rabi oscillation, i.e., $\tau_1\approx \pi/(2g_1)$, and be expressed as 
\begin{equation}
    \Delta^2 X_{\rm b}^{\rm min}\approx \frac{1}{2}e^{-2r}+\frac{\pi}{4}[ \frac{A_{\rm heat}}{g_1}+\frac{\Gamma}{2g_1}(1-e^{-2r})],
\end{equation}
where $A_{\rm heat}=\Gamma n_{\rm th}$ denotes the environment-induced decoherence rate. It can be seen that the squeezed state can be transferred from photons to acoustic phonons when the coupling strength overcomes the thermal reheating rate,
which is demonstrated by numerical simulations of $\Delta^2 X_{\rm b} (t)$ (solid curves) in Fig.~\ref{Fig3}(a). 
As the fidelity of two zero-mean Gaussian states are completely decided by their covariance matrices~\cite{adesso2008,chiribella2014}, the fidelity between initial signal photons with the state $\rho_{\rm sg}(0)$ and acoustic phonons with the state $\rho_{\rm b}(t)$, which quantifies the performance of the quantum protocol, can be calculated by $F[\rho_{\rm sg}(0),\rho_{\rm b}(t)]=F[\mathcal{V}_{\rm sg}(0),\mathcal{V}_{\rm b}(t)]$ (details see Appendix). We show numerical simulations of the fidelity $F[\rho_{\rm sg}(0),\rho_{\rm b}(t)]$ 
in the strong coupling regime in Fig.~\ref{Fig3}(a) (dash-dotted curves), revealing that high fidelity can be achieved at the optimal time $\tau_1$. 
Considering a short storage time $\tau_{\rm s}\ll 1/\Gamma$, the variance of the squeezed quadrature of retrieval photons can be approximately expressed as $\Delta^2 X_{\rm re}\approx \frac{1}{2}-\frac{1-e^{-2r}}{2}e^{-\frac{\Gamma}{2}t}\sin^2(g_2t) + \Gamma n_{\rm th}[ \frac{t}{2} + \frac{\pi}{4g_1} e^{-\frac{\Gamma}{2}t}\sin^2(g_2t) ]$, where $g_2$ corresponds to the effective optoacoustic coupling strength during the readout process.
With a Rabi oscillation, the minimum variance can be achieved at optimal time $\tau_2\approx\pi/(2g_2)$ and be well approximated by 
\begin{equation}
\Delta^2 X_{\rm re}^{\rm min} \approx\; \frac{1}{2}e^{-2r}+ \frac{\pi}{4}\!\left[
\frac{A_{\rm heat}}{g_1}
+ \frac{A_{\rm heat}}{g_2}
+ \frac{\Gamma}{2g_2}\left(1-e^{-2r}\right)
\right].
\end{equation}
It reveals that the squeezed state can be transferred from acoustic phonons to retrieval photons with a high efficiency when the optoacoustic coupling strength $g_2$ also exceeds the thermal reheating rate $A_{\rm heat}$. Such behavior can be demonstrated by numerical simulations of variance $\Delta^2 X_{\rm re} (t)$ (solid curves) and fidelity $\tilde{F}[\rho_{\rm sg}(0),\rho_{\rm re}(t)]$ (dash-dotted curves) as shown in Fig.~\ref{Fig3}(b), where $\rho_{\rm re}(t)$ is the density operator of retrieval photons.
We present the robustness of such protocol against environmental temperature in Fig.~\ref{Fig3}(c), showing that a high-fidelity memory of squeezed states is achievable in 
Brillouin-active waveguides at sub-Kelvin and higher temperatures.

We focus on a specific wave number $k$ in the above discussion. Since the optical and acoustic waves offer continuously accessible groups of photons and phonons in continuum optomechanical systems, this protocol enables the mapping of a squeezed state from signal photons onto retrieval photons over a broad bandwidth after a finite time storage in the acoustic phonons, as illustrated in Fig.~\ref{Fig3}(d). The squeezing factor is defined as the ratio between the squeezed quadrature variance $\Delta^2 X_{\rm re}$ and normalized quadrature variance of the vacuum noise (i.e., $0.5$)~\cite{mehmet2010}. 
To verify the quantumness of such a protocol, we consider an ensemble $\Lambda_{\rm s}$ of pure Gaussian squeezed vacuum states with random phase and 
unknown squeezing degree $r$, sampled from a Gaussian distribution $p_{\beta}(r)=\frac{1}{\beta\pi} e^{-\beta|r|^2}$. In this case, the quantum benchmark, i.e., the classical fidelity threshold amounts to the highest average fidelity achievable by means of measure-and-prepare strategies, is given by $\bar{\mathcal{F}}_{\rm c}=(1+\beta)/(2+\beta)$~\cite{adesso2008,chiribella2014}.
For $\beta=1$, the average fidelity $\tilde{F}$ after $50~{\rm ns}$ storage in acoustic phonons at $1~{\rm K}$ reaches $0.79$ when $g_1=g_2=60\Gamma$, exceeding the benchmark $\bar{\mathcal{F}}_{\rm c}\approx0.67$. This confirms that the quantumness of such pulsed protocol is therefore validated.
Realistically, such performance can be achieved with a writing pulsed pump of $7.8~{\rm ns}$ duration and $1.7~{\rm W}$ peak power, followed by a co-propagating readout
pulsed pump of $6.2~{\rm ns}$ duration and $2.7~{\rm W}$ peak power after $50~{\rm ns}$ of storage. Here, we assume a waveguide with a Brillouin gain $G_{\rm B}=750~{\rm m}^{-1}{\rm W}^{-1}$ and a length $L=6~{\rm cm}$ consistent with Ref.~\cite{morrison2017}. At an environmental temperature of $0.6~{\rm K}$, the average fidelity between idler and retrieval lights can reach $0.88$, exceeding the quantum benchmark $\bar{\mathcal{F}}_{\rm c}$. Further details on the quantum memory of squeezed thermal and squeezed coherent states are provided in the Appendix.\\

\noindent \textbf{Quantum storage of entangled states}.---After analyzing the quantum storage of
squeezed states in Brillouin-active waveguides, we now study the performance of the memory for entangled states.
We assume that a pair of entangled light waves are prepared at the initial time, i.e., an idler light $a_{\rm id}$ and
a signal light $a_{\rm sg}$ with the covariance matrix $\mathcal{V}_{\rm id,sg}(t=0)$. We feed one half of the entangled pair 
(idler light) to a single mode fiber for reference and the other half (signal light) to a Brillouin-active waveguide for memory. 
The memory protocol for entangled states is analogous to that for squeezed states. The state of the signal
light is first mapped onto acoustic phonons $b$ by applying a pulsed pump of duration $\tau_1$ during the writing process.
After a finite storage time $\tau_{s}$, the half of the entanlged pair stored on phonons is retrieved back onto a retrieval 
light $a_{\rm re}$ by utilizing a second pulsed pump with duration $\tau_2$ in the readout process. As we consider Gaussian
states for all optical and acoustic fields, we make use of the logarithmic negativity for quantifying bipartite entanglement.
The logarithmic negativity~\cite{plenio2005,vitali2007} between two systems is defined as $E_{\mathcal{N}} = {\rm max}\left[ 0, -{\rm ln} (2\lambda_{-}) \right]$,
where $\lambda_{-}$ is the minimal symplectic eigenvalue of the covariance matrix $\mathcal{V}$ between these two systems under 
a partial transposition and defined as $\lambda_{-}\equiv 2^{-1/2}\sqrt{ \Sigma(\mathcal{V}) - [ \Sigma(\mathcal{V})^2 - 4\det\mathcal{V} ]^{1/2} }$,
with $\Sigma(\mathcal{V}) \equiv \det A + \det B -2\det C$ (for more details see Appendix). The general criterion of entanglement for bimodal Gaussian states
requires the condition $E_{\mathcal{N}}>0$, which is equivalent to $\lambda_{-}<1/2$. 
The fidelity between different two-mode Gaussian states $\mathcal{V}_{1,2}$ with zero mean can be defined as
$F(\rho_1,\rho_2)=F(\mathcal{V}_1, \mathcal{V}_2)$, where $\mathcal{V}_{i}$ and $\rho_{i}$ ($i=1,2$) are 
the covariance matrix and density operator of each two-mode Gaussian state~\cite{banchi2015}.

%%%%%%%%%%%%%%%%%%%%%%%%%%%%%%%
%%%%%%%%%%%%%%%%%%%%%%%%%%%%%%%
\begin{figure}[t]
	\centerline{
	\includegraphics[clip, width=\linewidth]{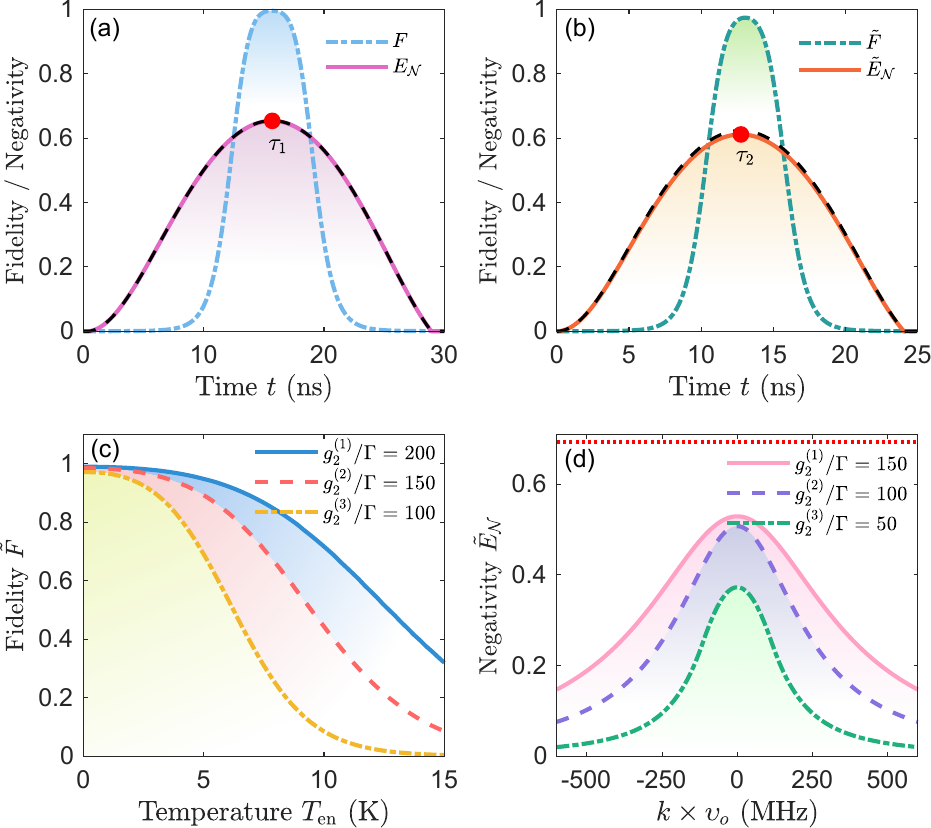}}
	\caption{ Time evolution of the negativity and fidelity in the strong coupling regime 
		during (a) the writing process and (b) the readout process (b), where red points denote the optimal
		time for the optimal quantum transduction and black dashed curves correspond to approximate
		analytical solutions of negativity. Parameters are $g_1/\Gamma=100$, $g_2/\Gamma=120$, $\tau_{\rm s}=5~{\rm ns}$, 
		and $T_{\rm en}=1~{\rm K}$. (c)  Variation of $\tilde{F}$ versus $T_{\rm en}$ for various ratios $g_2/\Gamma$,
		where $g_1=g_2$ and $\tau_{\rm s}=5~{\rm ns}$.
		(d) Continuum memory versus the wave number $k$ in the strong coupling regime ($g_1=g_2$) at $T_{\rm en}=1~{\rm K}$, 
		where red dotted line denotes the initial negativity between entangled idler and signal photons, and $\tau_{\rm s}=50~{\rm ns}$.
	}\label{Fig4}
\end{figure}
%%%%%%%%%%%%%%%%%%%%%%%%%%%%%%%
%%%%%%%%%%%%%%%%%%%%%%%%%%%%%%%

In the writing process, the minimal symplectic eigenvalues $\lambda_{-}$ of covariance matrix $\mathcal{V}_{\rm id,b}$
between idler photons $a_{\rm id}$ and acoustic phonons $b$ can be approximately expressed as
\begin{eqnarray}\label{Minimal symplectic eigenvalues in writing process}
	\lambda_{-} \approx \frac{1}{2}\times \frac{ 1 + 2n_{\rm th}\left( 1 - e^{-\frac{\Gamma}{2}t} \right) 
	+ \frac{\Gamma n_{\rm th}}{4g_1} e^{-\frac{\Gamma}{2}t} \sin(2g_1t) }
	{ 1 + e^{-\frac{\Gamma}{2}t} \sin^{2}(g_1t) }, \nonumber\\
\end{eqnarray}  
where $g_1$ denotes the effective optoacoustic coupling strength and the initial state of acoustic phonons 
is the ground state. The minimum value of $\lambda_{-}(t)$ can be obtained at the end of the first half Rabi oscillation cycle for a choice
$\tau_1\approx \pi/(2g_1)$ for which
\begin{equation}
E_{\mathcal{N}}^{\rm max} \approx -{\rm ln} \left[ \frac{1}{2} \left( 1 + \frac{\pi A_{\rm heat}}{2g_1} \right) \right].
\end{equation}
It indicates that the requirement for such a pulsed scheme to swap optical and acoustic states is that the optoacoustic coupling
strength overcomes the acoustic thermal reheating rate. We present the numerical simulations of the logarithmic negativity 
$E_{\mathcal{N}}$ (magenta solid curve) and fidelity $F\left [\mathcal{V}_{\rm id,sg}(0), \mathcal{V}_{\rm id,b}(t) \right ]$
(blue dash-dot curve) in Fig.~\ref{Fig4}(a), where the black 
dashed curve corresponds to the values evaluated by
Eq.~(\ref{Minimal symplectic eigenvalues in writing process}). After a short storage period $\tau_{\rm s}\ll 1/\Gamma$,
the minimal symplectic eigenvalues of covariance matrix $\mathcal{V}_{\rm id,re}$ between idler and retrieval photons in the
strong coupling regime can be given by 
\begin{eqnarray}\label{Minimal symplectic eigenvalues in readout process}
	&&\tilde{\lambda}_{-} \approx  \nonumber\\
	&& \frac{ 1 + 2n_{\rm th} \left( 1 - e^{-\frac{\Gamma}{2}t} \right)
	- \frac{\Gamma n_{\rm th}}{2} e^{-\frac{\Gamma}{2}t} 
	\left[ \frac{ \sin(2g_2t) }{g_2} - \frac{ \pi\sin^2(g_2t) }{g_1} \right]  }
	{ 2 \left[ 1 + e^{-\frac{\Gamma}{2}(t+\tau_1)} \sin^2(g_2t) \right]  }, \nonumber\\
\end{eqnarray}   
where $g_2$ denotes the effective optoacoustic coupling strength in the readout process and
$\tau_1$ corresponds to the optimal pulse duration of the first pump in the writing process.
The maximum entanglement between idler and retrieval photons can be achieved by utilizing
a second pulsed pump with the optimal duration $\tau_2\approx \pi/(2g_2)$ for which
%%%%%%%%%%%%%%%%%%%%%%%%%%%%%%
\begin{equation}
\tilde{E}_{\mathcal{N}}^{\rm max} \approx -{\rm ln} \left[ \frac{1}{2} \left( 1 + \frac{\pi A_{\rm heat}}{2g_2}
+ \frac{\pi A_{\rm heat}}{2g_1} e^{-\frac{\pi\Gamma}{4g_2}} \right) \right]
\end{equation}
%%%%%%%%%%%%%%%%%%%%%%%%%%%%%
The result shows that it is capable of retrieving one half of the entangled pair from acoustic phonons
when the effective coupling strengths $g_1$ and $g_2$ exceed the thermal reheating rate 
simultaneously, which is validated by numerical solutions of the logarithmic negativity
$\tilde{E}_{\mathcal{N}}$ (orange solid cure) and fidelity
$\tilde{F}\left[ \mathcal{V}_{\rm id,sg}(0), \mathcal{V}_{\rm id,re}(t) \right]$.

The robustness with $T_{\rm en}$ is 
illustrated in Fig.~\ref{Fig4}(c), which demonstrate that the pulsed protocol enables the storage 
of entangled states at sub-Kelvin and even higher temperatures.
Figure~\ref{Fig4}(d) presents the simulated dependence of negativity $\tilde{E}_{\mathcal{N}}$ 
on the wave number $k$ at a temperature of $1~$K, illustrating the ability of the system to swap 
states across a broad bandwidth of optical photons and acoustic phonons. The simulation shows that after 
$50~{\rm ns}$ of storage in acoustic phonons, one half of the entangled pair can be mapped onto
retrieval photons over a large interval of wave number $k$. Actually, if we apply a writing pump
with duration of $7.8~{\rm ns}$ and peak power of $1.7~{\rm W}$ to the waveguide with Brillouin gain $750~{\rm m}^{-1}{\rm W}^{-1}$ achieved in 
Ref.~\cite{morrison2017}, and then utilize a readout pump with duration of $5.2~{\rm ns}$ and peak power
of $3.8~{\rm W}$ after $30~{\rm ns}$ storage, the negativity between
idler and retrieval photons can reach $\tilde{E}_{\mathcal{N}}=0.59$ with a fidelity of 
$\tilde{F}=0.93$ at a temperature of $1~{\rm K}$.

\noindent \textbf{Conclusions and outlook}.---We have shown that quantum memory for light can be implemented in a continuous medium via Brillouin anti-Stokes scattering. The optomechanical interaction between optical and acoustic modes enables coherent photon–phonon state transfer in Brillouin-active waveguides, allowing the quantum state of a signal field to be stored in acoustic excitations and subsequently retrieved as photons. Owing to the strong photon–phonon coupling achievable in these systems, the protocol supports fast state swapping and high-fidelity storage over tens of nanoseconds, even in the presence of significant reheating.

Our analysis assumes a temperature-independent mechanical loss rate. In practice, mechanical dissipation decreases at lower temperatures, suggesting that substantially longer storage times should be achievable experimentally. For example, in state-of-the-art waveguides with a mechanical loss rate of $0.5\,\mathrm{MHz}$ at $T=0.8\,\mathrm{K}$, storage times for squeezed and entangled optical states can reach hundreds of nanoseconds. Although mechanical loss ultimately limits the storage duration, this constraint may be mitigated by employing crystalline materials, as recently demonstrated in Ref.~\cite{doeleman2023}. Importantly, because the protocol relies on coherent photon–phonon state swapping, ground-state cooling of the acoustic mode is not a prerequisite.

Experimentally, the retrieved quantum states can be characterized using standard homodyne detection of optical quadratures, while synchronization of the pump pulses required for storage of entangled states can be achieved with high-extinction modulators. Together with recent progress in continuum optomechanics—including cooling~\cite{otterstrom2018,zhu2023,johnson2023,blazquez2024,fischer2025brillouin}, strong coupling~\cite{martinez2025}, and photon–phonon entanglement~\cite{zhu2024}—the present scheme provides a realistic route toward broadband and scalable quantum memories based on optomechanical waveguides. These features position continuum optoacoustic platforms as promising candidates for future quantum communication and information-processing technologies.

\noindent \textbf{ Acknowledgements} The authors thank Liang Jiang for helpful discussions. This work is supported by the Max-Planck-Society through the independent
Max Planck Research Groups Scheme and the Deutsche Forschungsgemeinschaft (DFG, German Research Foundation) -- Project-ID 390833453 - EXC 2122 PhoenixD (``Photonics, Optics, and Engineering - Innovation across dimensions") -- Project-ID 429529648 -- TRR 306 QuCoLiMa (``Quantum Cooperativity of Light and Matter").

\bibliography{sample.bib}

@article{lvovsky2009,
  author={Lvovsky, Alexander I and Sanders, Barry C and Tittel, Wolfgang},
  title={Optical quantum memory},
  journal={Nat. Photonics},
  volume={3},
  number={12},
  pages={706},
  year={2009},
  publisher={Nature Publishing Group UK London},
  url = {https://doi.org/10.1038/nphoton.2009.231}
}

@article{afzelius2015,
  title={Quantum memory for photons},
  author={Afzelius, Mikael and Gisin, Nicolas and De Riedmatten, Hugues},
  journal={Phys. Today},
  volume={68},
  number={12},
  pages={42},
  year={2015},
  publisher={AIP Publishing},
  url = {https://doi.org/10.1063/PT.3.3021}
}

@article{kimble2008,
  title={The quantum internet},
  author={Kimble, H Jeff},
  journal={Nature},
  volume={453},
  number={7198},
  pages={1023},
  year={2008},
  publisher={Nature Publishing Group},
  url = {https://doi.org/10.1038/nature07127}
}

@article{wehner2018,
  title={Quantum internet: A vision for the road ahead},
  author={Wehner, Stephanie and Elkouss, David and Hanson, Ronald},
  journal={Science},
  volume={362},
  number={6412},
  year={2018},
  publisher={American Association for the Advancement of Science},
  url = {https://doi.org/10.1126/science.aam9288}
}

@article{mazelanik2021,
  title={Real-time ghost imaging of Bell-nonlocal entanglement between a photon and a quantum memory},
  author={Mazelanik, Mateusz and Leszczy{\'n}ski, Adam and Lipka, Micha{\l} and Wasilewski, Wojciech and Parniak, Micha{\l}},
  journal={Quantum},
  volume={5},
  pages={493},
  year={2021},
  publisher={Verein zur F{\"o}rderung des Open Access Publizierens in den Quantenwissenschaften},
  url = {https://doi.org/10.22331/q-2021-07-01-493}
}

@article{zaiser2016,
  title={Enhancing quantum sensing sensitivity by a quantum memory},
  author={Zaiser, Sebastian and Rendler, Torsten and Jakobi, Ingmar and Wolf, Thomas and Lee, Sang-Yun and Wagner, Samuel and Bergholm, Ville and Schulte-Herbr{\"u}ggen, Thomas and Neumann, Philipp and Wrachtrup, J{\"o}rg},
  journal={Nat. Commun.},
  volume={7},
  number={1},
  pages={12279},
  year={2016},
  publisher={Nature Publishing Group UK London},
  url = {https://doi.org/10.1038/ncomms12279}
}

@article{degen2017,
  title={Quantum sensing},
  author={Degen, Christian L and Reinhard, Friedemann and Cappellaro, Paola},
  journal={Rev. Mod. Phys.},
  volume={89},
  number={3},
  pages={035002},
  year={2017},
  publisher={APS},
  url = {https://doi.org/10.1103/RevModPhys.89.035002}
}

@article{mol2023,
  title={Quantum memories for fundamental science in space},
  author={Mol, Jan-Michael and Esguerra, Luisa and Meister, Matthias and Bruschi, David Edward and Schell, Andreas Wolfgang and Wolters, Janik and W{\"o}rner, Lisa},
  journal={Quantum Sci. Technol.},
  volume={8},
  number={2},
  pages={024006},
  year={2023},
  publisher={IOP Publishing},
  url = {https://doi.org/10.1088/2058-9565/acb2f1}
}

@article{heshami2016,
  title={Quantum memories: emerging applications and recent advances},
  author={Heshami, Khabat and England, Duncan G and Humphreys, Peter C and Bustard, Philip J and Acosta, Victor M and Nunn, Joshua and Sussman, Benjamin J},
  journal={J. Mod. Opt.},
  volume={63},
  number={20},
  pages={2005},
  year={2016},
  publisher={Taylor \& Francis},
  url = {https://doi.org/10.1080/09500340.2016.1148212}
}

@article{lei2023,
  title={Quantum optical memory for entanglement distribution},
  author={Lei, Yisheng and Kimiaee Asadi, Faezeh and Zhong, Tian and Kuzmich, Alex and Simon, Christoph and Hosseini, Mahdi},
  journal={Optica},
  volume={10},
  number={11},
  pages={1511},
  year={2023},
  publisher={Optica Publishing Group},
  url = {https://doi.org/10.1364/OPTICA.493732}
}

@article{julsgaard2004,
  title={Experimental demonstration of quantum memory for light},
  author={Julsgaard, Brian and Sherson, Jacob and Cirac, J Ignacio and Fiur{\'a}{\v{s}}ek, Jarom{\'\i}r and Polzik, Eugene S},
  journal={Nature},
  volume={432},
  number={7016},
  pages={482},
  year={2004},
  publisher={Nature Publishing Group UK London},
  url = {https://doi.org/10.1038/nature03064}
}

@article{li2016,
  title={Quantum memory with strong and controllable Rydberg-level interactions},
  author={Li, Lin and Kuzmich, A},
  journal={Nat. Commun.},
  volume={7},
  number={1},
  pages={13618},
  year={2016},
  publisher={Nature Publishing Group UK London},
  url = {https://doi.org/10.1038/ncomms13618}
}

@article{clausen2012,
  title={Quantum Storage of Heralded Polarization Qubits in Birefringent and Anisotropically<? format?> Absorbing Materials},
  author={Clausen, Christoph and Bussieres, F{\'e}lix and Afzelius, Mikael and Gisin, Nicolas},
  journal={Phy. Rev. Lett.},
  volume={108},
  number={19},
  pages={190503},
  year={2012},
  publisher={APS},
  url = {https://doi.org/10.1103/PhysRevLett.108.190503}
}

@article{gundougan2012,
  title={Quantum storage of a photonic polarization qubit in a solid},
  author={G{\"u}ndo{\u{g}}an, Mustafa and Ledingham, Patrick M and Almasi, Attaallah and Cristiani, Matteo and De Riedmatten, Hugues},
  journal={Phy. Rev. Lett.},
  volume={108},
  number={19},
  pages={190504},
  year={2012},
  publisher={APS},
  url = {https://doi.org/10.1103/PhysRevLett.108.190504}
}

@article{zhou2012,
  title={Realization of reliable solid-state quantum memory for photonic polarization qubit},
  author={Zhou, Zong-Quan and Lin, Wei-Bin and Yang, Ming and Li, Chuan-Feng and Guo, Guang-Can},
  journal={Phys. Rev. Lett.},
  volume={108},
  number={19},
  pages={190505},
  year={2012},
  publisher={APS},
  url = {https://doi.org/10.1103/PhysRevLett.108.190505}
}

@article{shandilya2021,
  title={Optomechanical interface between telecom photons and spin quantum memory},
  author={Shandilya, Prasoon K and Lake, David P and Mitchell, Matthew J and Sukachev, Denis D and Barclay, Paul E},
  journal={Nat. Phys.},
  volume={17},
  number={12},
  pages={1420},
  year={2021},
  publisher={Nature Publishing Group UK London},
  url = {https://doi.org/10.1038/s41567-021-01364-3}
}

@article{wallucks2020,
  title={A quantum memory at telecom wavelengths},
  author={Wallucks, Andreas and Marinkovi{\'c}, Igor and Hensen, Bas and Stockill, Robert and Gr{\"o}blacher, Simon},
  journal={Nat. Phys.},
  volume={16},
  number={7},
  pages={772},
  year={2020},
  publisher={Nature Publishing Group UK London},
  url = {https://doi.org/10.1038/s41567-020-0891-z}
}

@article{aspelmeyer2014,
  title={Cavity optomechanics},
  author={Aspelmeyer, Markus and Kippenberg, Tobias J and Marquardt, Florian},
  journal={Rev. Mod. Phys.},
  volume={86},
  number={4},
  pages={1391},
  year={2014},
  publisher={APS},
  url = {https://doi.org/10.1103/RevModPhys.86.1391}
}

@article{metcalfe2014,
  title={Applications of cavity optomechanics},
  author={Metcalfe, Michael},
  journal={Appl. Phys. Rev.},
  volume={1},
  number={3},
  pages={031105},
  year={2014},
  publisher={AIP Publishing},
  url = {https://doi.org/10.1063/1.4896029}
}

@article{barzanjeh2022,
  title={Optomechanics for quantum technologies},
  author={Barzanjeh, Shabir and Xuereb, Andr{\'e} and Gr{\"o}blacher, Simon and Paternostro, Mauro and Regal, Cindy A and Weig, Eva M},
  journal={Nat. Phys.},
  volume={18},
  number={1},
  pages={15},
  year={2022},
  publisher={Nature Publishing Group UK London},
  url = {https://doi.org/10.1038/s41567-021-01402-0}
}

@article{van2016,
  title={Unifying Brillouin scattering and cavity optomechanics},
  author={Van Laer, Rapha{\"e}l and Baets, Roel and Van Thourhout, Dries},
  journal={Phys. Rev. A},
  volume={93},
  number={5},
  pages={053828},
  year={2016},
  publisher={APS},
  url = {https://doi.org/10.1103/PhysRevA.93.053828}
}

@article{rakich2018,
  title={Quantum theory of continuum optomechanics},
  author={Rakich, Peter and Marquardt, Florian},
  journal={New Journal of Physics},
  volume={20},
  number={4},
  pages={045005},
  year={2018},
  publisher={IOP Publishing},
  url = {https://doi.org/10.1088/1367-2630/aaac4f}
}

@article{zoubi2017,
  title={Quantum Nonlinear Optics in Optomechanical Nanoscale Waveguides.},
  author={Zoubi, H and Hammerer, K},
  journal={Phys. Rev. Lett.},
  volume={119},
  number={12},
  pages={123602},
  year={2017},
  url = {https://doi.org/10.1103/PhysRevLett.119.123602}
}

@article{otterstrom2018,
  title={Optomechanical cooling in a continuous system},
  author={Otterstrom, Nils T and Behunin, Ryan O and Kittlaus, Eric A and Rakich, Peter T},
  journal={Phys. Rev. X},
  volume={8},
  number={4},
  pages={041034},
  year={2018},
  publisher={APS},
  url = {https://doi.org/10.1103/PhysRevX.8.041034}
}

@article{zhang2023,
  title={Quantum coherent control in pulsed waveguide optomechanics},
  author={Zhang, Junyin and Zhu, Chang Long and Wolff, Christian and Stiller, Birgit},
  journal={Phys. Rev. Research},
  volume={5},
  number={1},
  pages={013010},
  year={2023},
  publisher={APS},
  url = {https://doi.org/10.1103/PhysRevResearch.5.013010}
}

@article{zhu2023,
  title={Dynamic Brillouin cooling for continuous optomechanical systems},
  author={Zhu, Chang Long and Stiller, Birgit},
  journal={Mater. Quantum Technol.},
  volume={3},
  number={1},
  pages={015003},
  year={2023},
  publisher={IOP Publishing},
  url = {https://doi.org/10.1088/2633-4356/acc2a5}
}

@article{johnson2023,
  title={Laser Cooling of Traveling-Wave Phonons in an Optical Fiber},
  author={Johnson, Joel N and Haverkamp, Danielle R and Ou, Yi-Hsin and Kieu, Khanh and Otterstrom, Nils T and Rakich, Peter T and Behunin, Ryan O},
  journal={Phys. Rev. Appl.},
  volume={20},
  number={3},
  pages={034047},
  year={2023},
  publisher={APS},
  url = {https://doi.org/10.1103/PhysRevApplied.20.034047}
}

@article{blazquez2024,
  title={Optoacoustic cooling of traveling hypersound waves},
  author={Bl{\'a}zquez Mart{\'\i}nez, Laura and Wiedemann, Philipp and Zhu, Chang Long and Geilen, Andreas and Stiller, Birgit},
  journal={Phys. Rev. Lett.},
  volume={132},
  number={2},
  pages={023603},
  year={2024},
  publisher={APS},
  url = {https://doi.org/10.1103/PhysRevLett.132.023603}
}

@article{zhu2024,
  title={Optoacoustic entanglement in a continuous Brillouin-active solid state system},
  author={Zhu, Chang Long and Genes, Claudiu and Stiller, Birgit},
  journal={Phys. Rev. Lett.},
  volume={133},
  number={20},
  pages={203602},
  year={2024},
  publisher={APS},
  url = {https://doi.org/10.1103/PhysRevLett.133.203602}
}

@article{zoubi2024,
  title={Slow light through Brillouin scattering in continuum quantum optomechanics},
  author={Zoubi, Hashem and Hammerer, Klemens},
  journal={Front. Quantum Sci. Technol.},
  volume={3},
  pages={1437933},
  year={2024},
  publisher={Frontiers Media SA},
  url = {https://doi.org/10.3389/frqst.2024.1437933}
}

@article{cryer2025,
  title={Brillouin--Mandelstam scattering in telecommunications optical fiber at millikelvin temperatures},
  author={Cryer-Jenkins, EA and Leung, AC and Rathee, H and Tan, AKC and Major, KD and Vanner, MR},
  journal={APL Photonics},
  volume={10},
  pages = {010805},
  year={2025},
  publisher={AIP Publishing},
  url = {https://doi.org/10.1063/5.0241253}
}

@article{stiller2019,
  title={Cross talk-free coherent multi-wavelength Brillouin interaction},
  author={Stiller, Birgit and Merklein, Moritz and Vu, Khu and Ma, Pan and Madden, Stephen J and Poulton, Christopher G and Eggleton, Benjamin J},
  journal={APL Photonics},
  volume={4},
  pages = {040802},
  number={4},
  year={2019},
  publisher={AIP Publishing},
  url = {https://doi.org/10.1063/1.5087180}
}

@article{becker2024,
  title={An optoacoustic field-programmable perceptron for recurrent neural networks},
  author={Becker, Steven and Englund, Dirk and Stiller, Birgit},
  journal={Nat. Commun.},
  volume={15},
  number={1},
  pages={3020},
  year={2024},
  publisher={Nature Publishing Group UK London},
  url = {https://doi.org/10.1038/s41467-024-47053-6}
}

@article{eggleton2019,
  title={Brillouin integrated photonics},
  author={Eggleton, Benjamin J and Poulton, Christopher G and Rakich, Peter T and Steel, Michael J and Bahl, Gaurav},
  journal={Nat. Photonics},
  volume={13},
  number={10},
  pages={664},
  year={2019},
  publisher={Nature Publishing Group UK London},
  url = {https://doi.org/10.1038/s41566-019-0498-z}
}

@article{gyger2020,
  title={Observation of stimulated Brillouin scattering in silicon nitride integrated waveguides},
  author={Gyger, Flavien and Liu, Junqiu and Yang, Fan and He, Jijun and Raja, Arslan S and Wang, Rui Ning and Bhave, Sunil A and Kippenberg, Tobias J and Th{\'e}venaz, Luc},
  journal={Phys. Rev. Lett.},
  volume={124},
  number={1},
  pages={013902},
  year={2020},
  publisher={APS},
  url ={https://doi.org/10.1103/PhysRevLett.124.013902}
}

@article{botter2022,
  title={Guided-acoustic stimulated Brillouin scattering in silicon nitride photonic circuits},
  author={Botter, Roel and Ye, Kaixuan and Klaver, Yvan and Suryadharma, Radius and Daulay, Okky and Liu, Gaojian and van den Hoogen, Jasper and Kanger, Lou and van der Slot, Peter and Klein, Edwin and others},
  journal={Sci. Adv.},
  volume={8},
  number={40},
  pages={eabq2196},
  year={2022},
  publisher={American Association for the Advancement of Science},
  url = {https://doi.org/10.1126/sciadv.abq2196}
}

@article{rodrigues2025,
  title={Cross-polarized stimulated Brillouin scattering in lithium niobate waveguides},
  author={Rodrigues, Caique C and Schilder, Nick J and Zurita, Roberto O and Magalh{\~a}es, Let{\'\i}cia S and Shams-Ansari, Amirhassan and Dos Santos, Felipe JL and Paiano, Ot{\'a}vio M and Alegre, Thiago PM and Lon{\v{c}}ar, Marko and Wiederhecker, Gustavo S},
  journal={Phys. Rev. Lett.},
  volume={134},
  number={11},
  pages={113601},
  year={2025},
  publisher={APS},
  url = {https://doi.org/10.1103/PhysRevLett.134.113601}
}

@article{martinez2025,
  title={Cavity-less Brillouin strong coupling in a solid-state continuous system},
  author={Mart{\'\i}nez, Laura Bl{\'a}zquez and Zhu, Chang Long and Stiller, Birgit},
  journal={arXiv preprint arXiv:2507.08673},
  year={2025},
  url = {https://doi.org/10.48550/arXiv.2507.08673}
}

@article{kharel2016,
  title={Noise and dynamics in forward Brillouin interactions},
  author={Kharel, Prashanta and Behunin, Ryan O and Renninger, William H and Rakich, Peter T},
  journal={Phys. Rev. A},
  volume={93},
  number={6},
  pages={063806},
  year={2016},
  publisher={APS},
  url = {https://doi.org/10.1103/PhysRevA.93.063806}
}

@article{sipe2016,
  title={A Hamiltonian treatment of stimulated Brillouin scattering in nanoscale integrated waveguides},
  author={Sipe, JE and Steel, MJ},
  journal={New J. Phys.},
  volume={18},
  number={4},
  pages={045004},
  year={2016},
  publisher={IOP Publishing},
  url = {https://doi.org/10.1088/1367-2630/18/4/045004}
}

@article{zoubi2016,
  title={Optomechanical multimode Hamiltonian for nanophotonic waveguides},
  author={Zoubi, Hashem and Hammerer, Klemens},
  journal={Phys. Rev. A},
  volume={94},
  number={5},
  pages={053827},
  year={2016},
  publisher={APS},
  url = {https://doi.org/10.1103/PhysRevA.94.053827}
}

@article{morrison2017,
  title={Compact Brillouin devices through hybrid integration on silicon},
  author={Morrison, Blair and Casas-Bedoya, Alvaro and Ren, Guanghui and Vu, Khu and Liu, Yang and Zarifi, Atiyeh and Nguyen, Thach G and Choi, Duk-Yong and Marpaung, David and Madden, Stephen J and others},
  journal={Optica},
  volume={4},
  number={8},
  pages={847},
  year={2017},
  publisher={Optical Society of America},
  url = {https://doi.org/10.1364/OPTICA.4.000847}
}

@article{wang2022,
  title={Generalized matching condition for unity efficiency quantum transduction},
  author={Wang, Chiao-Hsuan and Zhang, Mengzhen and Jiang, Liang},
  journal={Phys. Rev. Research},
  volume={4},
  number={4},
  pages={L042023},
  year={2022},
  publisher={APS},
  url = {https://doi.org/10.1103/PhysRevResearch.4.L042023}
}

@article{lauk2020,
  title={Perspectives on quantum transduction},
  author={Lauk, Nikolai and Sinclair, Neil and Barzanjeh, Shabir and Covey, Jacob P and Saffman, Mark and Spiropulu, Maria and Simon, Christoph},
  journal={Quantum Sci. Technol.},
  volume={5},
  number={2},
  pages={020501},
  year={2020},
  publisher={IOP Publishing},
  url = {https://doi.org/10.1088/2058-9565/ab788a}
}

@article{andersen2015,
  title={Hybrid discrete-and continuous-variable quantum information},
  author={Andersen, Ulrik L and Neergaard-Nielsen, Jonas S and Van Loock, Peter and Furusawa, Akira},
  journal={Nat. Phys.},
  volume={11},
  number={9},
  pages={713},
  year={2015},
  publisher={Nature Publishing Group UK London},
  url = {https://doi.org/10.1038/nphys3410}
}

@article{ye2024,
  title={Essay: Quantum sensing with atomic, molecular, and optical platforms for fundamental physics},
  author={Ye, Jun and Zoller, Peter},
  journal={Phys. Rev. Lett.},
  volume={132},
  number={19},
  pages={190001},
  year={2024},
  publisher={APS},
  url = {https://doi.org/10.1103/PhysRevLett.132.190001}
}

@article{weedbrook2012,
  title={Gaussian quantum information},
  author={Weedbrook, Christian and Pirandola, Stefano and Garc{\'\i}a-Patr{\'o}n, Ra{\'u}l and Cerf, Nicolas J and Ralph, Timothy C and Shapiro, Jeffrey H and Lloyd, Seth},
  journal={Rev. Mod. Phys.},
  volume={84},
  number={2},
  pages={621},
  year={2012},
  publisher={APS},
  url = {https://doi.org/10.1103/RevModPhys.84.621}
}

@article{adesso2008,
  title={Quantum benchmark for teleportation and storage of squeezed states},
  author={Adesso, Gerardo and Chiribella, Giulio},
  journal={Phys. Rev. Lett.},
  volume={100},
  number={17},
  pages={170503},
  year={2008},
  publisher={APS},
  url = {https://doi.org/10.1103/PhysRevLett.100.170503}
}

@article{chiribella2014,
  title={Quantum benchmarks for pure single-mode Gaussian states},
  author={Chiribella, Giulio and Adesso, Gerardo},
  journal={Phys. Rev. Lett.},
  volume={112},
  number={1},
  pages={010501},
  year={2014},
  publisher={APS},
  url = {https://doi.org/10.1103/PhysRevLett.112.010501}
}

@article{mehmet2010,
  title={Observation of squeezed states with strong photon-number oscillations},
  author={Mehmet, Moritz and Vahlbruch, Henning and Lastzka, Nico and Danzmann, Karsten and Schnabel, Roman},
  journal={Phys. Rev. A},
  volume={81},
  number={1},
  pages={013814},
  year={2010},
  publisher={APS},
  url = {https://doi.org/10.1103/PhysRevA.81.013814}
}

@article{plenio2005,
  title={Logarithmic negativity: a full entanglement monotone that is not convex},
  author={Plenio, Martin B},
  journal={Phys. Rev. Lett.},
  volume={95},
  number={9},
  pages={090503},
  year={2005},
  publisher={APS},
  url = {https://doi.org/10.1103/PhysRevLett.95.090503}
}

@article{vitali2007,
  title={Optomechanical entanglement between a movable mirror and a cavity field},
  author={Vitali, David and Gigan, Sylvain and Ferreira, Anderson and B{\"o}hm, HR and Tombesi, Paolo and Guerreiro, Ariel and Vedral, Vlatko and Zeilinger, <? format?> A and Aspelmeyer, Markus},
  journal={Phys. Rev. Lett.},
  volume={98},
  number={3},
  pages={030405},
  year={2007},
  publisher={APS},
  url = {https://doi.org/10.1103/PhysRevLett.98.030405}
}

@article{banchi2015,
  title={Quantum fidelity for arbitrary Gaussian states},
  author={Banchi, Leonardo and Braunstein, Samuel L and Pirandola, Stefano},
  journal={Phys. Rev. Lett.},
  volume={115},
  number={26},
  pages={260501},
  year={2015},
  publisher={APS},
  url = {https://doi.org/10.1103/PhysRevLett.115.260501}
}

@article{doeleman2023,
  title={Brillouin optomechanics in the quantum ground state},
  author={Doeleman, Hugo Michiel and Schatteburg, Tom and Benevides, R and Vollenweider, S and Macri, Dorotea and Chu, Yiwen},
  journal={Phys. Rev. Research},
  volume={5},
  number={4},
  pages={043140},
  year={2023},
  publisher={APS},
  url = {https://doi.org/10.1103/PhysRevResearch.5.043140}
}

@article{fischer2025brillouin,
  title={Brillouin--Mandelstam scattering-based cooling of traveling acoustic waves from cryogenic temperatures},
  author={Fischer, Lisa and Bl{\'a}zquez Mart{\'\i}nez, Laura and Zhu, Chang Long and Chenevi{\`e}re, Robin and Troles, Johann and Stiller, Birgit},
  journal={Optics Letters},
  volume={51},
  number={1},
  pages={121--124},
  year={2025},
  publisher={Optica Publishing Group},
   url = {https://doi.org/10.1364/OL.581872}
}

@article{liu2023coherent,
  title={Coherent memory for microwave photons based on long-lived mechanical excitations},
  author={Liu, Yulong and Liu, Qichun and Sun, Huanying and Chen, Mo and Wang, Shuaipeng and Li, Tiefu},
  journal={NPJ Quantum Information},
  volume={9},
  number={1},
  pages={80},
  year={2023},
  publisher={Nature Publishing Group UK London},
  url = {https://doi.org/10.1038/s41534-023-00749-x}
}

\newpage
\onecolumngrid

\appendix

\section{Quantum transduction via Brillouin anti-Stokes scattering}\label{S1}
We consider the backward Brillouin anti-Stokes scattering in an optical waveguide, which offers an
optomechanical interaction between a pump light $A_{\rm p}$, a scattered light $A_{\rm as}$, and an acoustic field $B_{\rm ac}$. 
The dynamics of this Brillouin optomechanics can be given by~\cite{kharel2016,sipe2016,zoubi2016}
\begin{eqnarray}\label{Dynamics of Brillouin optomechanics in waveguides}
	\frac{\partial A_{\rm p}}{\partial t} + \upsilon_{\rm o} \frac{\partial A_{\rm p}}{\partial z} &=& 
	- \frac{\gamma}{2} A_{\rm p} - i g_0 A_{\rm as} B_{\rm ac}^{\dagger} + \sqrt{\gamma} \tilde{\xi}_{\rm p}, \nonumber\\
	\frac{\partial A_{\rm as}}{\partial t} - \upsilon_{\rm o} \frac{\partial A_{\rm as}}{\partial z} &=&
	- \frac{\gamma}{2} A_{\rm as} - i g_0 A_{\rm p} B_{\rm ac} + \sqrt{\gamma} \tilde{\xi}_{\rm as}, \nonumber\\
	\frac{\partial B_{\rm ac}}{\partial t} - \upsilon_{\rm ac} \frac{\partial B_{\rm ac}}{\partial z} &=&
	- \frac{\Gamma}{2} B_{\rm ac} - i g_0 A_{\rm p}^{\dagger} A_{\rm as} + \sqrt{\Gamma} \tilde{\xi}_{\rm ac},
\end{eqnarray}
where $A_{\rm p}$, $A_{\rm as}$, and $B_{ac}$ denote envelope operators of the pump field, anti-Stokes field,
and acoustic field, respectively. $\gamma$ ($\Gamma$) and $\upsilon_{\rm o}$ ($\upsilon_{\rm ac}$) represent optical 
(acoustic) damping rate and group velocity. $g_0$ is the optomechanical coupling strength at the single quanta level,
where we take it real and positive with loss of generality. $\tilde{\xi}_{\rm p}$, $\tilde{\xi}_{\rm as}$, and
$\tilde{\xi}_{\rm ac}$ correspond to Langevin noises for these three fields and obey the following statistical properties~\cite{zhu2024}
\begin{eqnarray}\label{Noises properties}
	\langle \tilde{\xi}_\text{p}(t,z)\rangle &=& \langle \tilde{\xi}_\text{as}(t,z) = \langle \tilde{\xi}_\text{ac}(t,z)= 0, \nonumber\\
	\langle \tilde{\xi}_\text{p}^{\dagger} (t_1,z_1) \tilde{\xi}_\text{p}(t_2,z_2)\rangle &=&
	\langle \tilde{\xi}_\text{as}^{\dagger}(t_1,z_1) \tilde{\xi}_\text{as}(t_2,z_2)\rangle = 0, \nonumber\\
	\langle \tilde{\xi}_\text{ac}^{\dagger}(t_1,z_1) \tilde{\xi}_\text{ac}(t_2,z_2)\rangle &=& n_\text{th}\delta(t_1-t_2)\delta(z_1-z_2),	
\end{eqnarray}
where $n_\text{th}=1/(e^{\hbar\Omega_\text{ac}/k_\text{B} T_\text{en}}-1)$ is the thermal phonon occupation
of the acoustic field at temperature $T_\text{en}$, $\Omega_\text{ac}$ is the acoustic frequency, and $k_\text{B}$
is Boltzmann constant. Considering an undepleted pump, i.e., undepleted-pump approximation, the three-wave interaction
can be reduced to an effective interaction between the anti-Stokes and acoustic waves with motion equations
\begin{eqnarray}
	\frac{\partial A_{\rm as}}{\partial t} - \upsilon_{\rm o} \frac{\partial A_{\rm as}}{\partial z} &=&
	- \frac{\gamma}{2} A_{\rm as} - i g B_{\rm ac} + \sqrt{\gamma} \tilde{\xi}_{\rm as}, \nonumber\\
	\frac{\partial B_{\rm ac}}{\partial t} - \upsilon_{\rm ac} \frac{\partial B_{\rm ac}}{\partial z} &=&
	- \frac{\Gamma}{2} B_{\rm ac} - i g A_{\rm as} + \sqrt{\Gamma} \tilde{\xi}_{\rm ac},	
\end{eqnarray}
where $g=g_0\sqrt{A^{\dagger}_{\rm p} A_{\rm p}}$ denotes the effective coupling strength between anti-Stokes
photons and acoustic phonons, which is enhanced by the pump wave. In the discrete variable representation, the 
anti-Stokes and acoustic envelope operators can be given by
\begin{eqnarray}\label{Discrete picture of reduced interacton}
	A_\text{as}(t,z) &=& \frac{1}{\sqrt{2\pi}} \int a(t,k) e^{ikz}d k,\nonumber\\
	B_\text{ac}(t,z) &=& \frac{1}{\sqrt{2\pi}} \int b(t,k) e^{ikz}d k.
\end{eqnarray}
Therefore, the motion equations of the reduced Brillouin anti-Stokes process in the momentum space can be given by
\begin{eqnarray}
	\frac{d a(t,k)}{d t} &=& \left( -\frac{\gamma}{2} + i\Delta_{\rm as} \right) a(t,k) - ig b(t,k) + \sqrt{\gamma} \xi_{\rm as}, \nonumber\\
	\frac{d b(t,k)}{d t} &=& \left( -\frac{\Gamma}{2} + i\Delta_{\rm ac} \right) b(t,k) - ig a(t,k) + \sqrt{\Gamma} \xi_{\rm ac}.
\end{eqnarray}
Here, $a(t,k)$ ($b(t,k)$) denotes the annihilation operator for the $k_{\rm th}$ anti-Stokes (acoustic) mode and
$\xi_{\rm as}(t,k)$ ($\xi_{\rm ac}(t,k)$) corresponds to the Fourier transformation of the optical (acoustic)
Langevin noise $\tilde{\xi}_{\rm as}(t,z)$ ($\tilde{\xi}_{\rm ac}(t,z)$). Based on the statistical properties
of Langevin noises described in Eq.~(\ref{Noises properties}), the dynamics of the mean phonon number and mean 
photon number can be expressed as
%%%%%%%%%%%%%%%%%%%%%%%%%%%%%%%%%%%%%%%%%%%%%%%%%%%%%%%%%%%%%%
\begin{eqnarray}
	\frac{d n_a}{dt} &=& -\gamma n_a - ig \left( \langle a^{\dagger}b \rangle + \langle a^{\dagger}b \rangle^*  \right), \nonumber\\
	\frac{d n_b}{dt} &=& -\Gamma n_b + ig \left( \langle a^{\dagger}b \rangle + \langle a^{\dagger}b \rangle^*  \right)
	+ \Gamma n_{\rm th}, \nonumber\\
	\frac{d \langle a^{\dagger}b \rangle }{dt} &=& - \left( i(\Delta_{\rm as} - \Delta_{\rm ac})
	+ \frac{\gamma+\Gamma}{2} \right) \langle a^{\dagger}b \rangle - ig n_a + ig n_b.
\end{eqnarray}
%%%%%%%%%%%%%%%%%%%%%%%%%%%%%%%%%%%%%%%%%%%%%%%%%%%%%%%%%%%%%%
$n_a = \langle a^{\dagger} a \rangle$ and $n_b = \langle b^{\dagger} b \rangle$ denote the operators of
the mean photon and mean phonon number, respectively. In the strong coupling regime, i.e., $g\gg \gamma, \Gamma$, the analytical solutions of the mean
photon number and phonon number can be written as follows
%%%%%%%%%%%%%%%%%%%%%%%%%%%%%%%%%%%%%%%%%%%%%%%%%%%%%%%%%%%%%%
\begin{eqnarray}
	n_a(k,t) &=& A_1 e^{-\frac{\gamma+\Gamma}{2}t}	+ A_2 e^{-\frac{\gamma+\Gamma}{2}t}\cos(\Omega t)
	+ A_3 e^{-\frac{\gamma+\Gamma}{2}t} \sin(\Omega t) + n_a^{\rm ss}, \\
	n_b(k,t) &=& B_1 e^{-\frac{\gamma+\Gamma}{2}t}  + B_2 e^{-\frac{\gamma+\Gamma}{2}t}\cos(\Omega t)
	+ B_3 e^{-\frac{\gamma+\Gamma}{2}t} \sin(\Omega t) + n_b^{\rm ss}.
\end{eqnarray}
%%%%%%%%%%%%%%%%%%%%%%%%%%%%%%%%%%%%%%%%%%%%%%%%%%%%%%%%%%%%%%
with coefficients
%%%%%%%%%%%%%%%%%%%%%%%%%%%%%%%%%%%%%%%%%%%%%%%%%%%%%%%%%%%%%%
\begin{eqnarray}
	A_1 &=& \frac{ -8g^2 + (\Gamma-\gamma)^2 + 4\Omega^2 }{4\Omega^2}n_{a,0}
	+ \frac{2g^2}{\Omega^2} n_{b,0} - \frac{ (\Gamma+\gamma)^2+4\Omega^2 }{4\Omega^2}n_{a}^{\rm ss}, \nonumber\\
	A_2 &=& \frac{ 8g^2-(\Gamma-\gamma)^2 }{ 4\Omega^2 } n_{a,0} - \frac{2g^2}{\Omega^2} n_{b,0}
	+ \frac{(\Gamma+\gamma)^2}{4\Omega^2} n_{a}^{\rm ss}, \nonumber\\
	A_3 &=& \frac{\Gamma-\gamma}{2\Omega} n_{a,0} - \frac{\Gamma+\gamma}{2\Omega} n_{a}^{\rm ss}, \nonumber\\
	B_1 &=& \frac{2g^2}{\Omega^2} n_{a,0} + \frac{ -8g^2+(\Gamma-\gamma)^2+4\Omega^2 }{4\Omega^2} n_{b,0}
	- \frac{ (\Gamma+\gamma)^2+4\Omega^2 }{4\Omega^2}n_{b}^{\rm ss} + \frac{\gamma\Gamma}{\Omega^2} n_{\rm th}, \nonumber\\
	B_2 &=& -\frac{2g^2}{\Omega^2} n_{a,0} + \frac{ 8g^2-(\Gamma-\gamma)^2 }{4\Omega^2} n_{b,0}
	+ \frac{(\Gamma+\gamma)^2}{4\Omega^2} n_{b}^{\rm ss} - \frac{\gamma\Gamma}{\Omega^2} n_{\rm th}, \nonumber\\
	B_3 &=& -\frac{\Gamma-\gamma}{2\Omega} n_{b,0} + \frac{\Gamma}{\Omega} n_{\rm th} - \frac{\Gamma+\gamma}{2\Omega} n_{b}^{\rm ss},
\end{eqnarray}
%%%%%%%%%%%%%%%%%%%%%%%%%%%%%%%%%%%%%%%%%%%%%%%%%%%%%%%%%%%%%%
and steady states
%%%%%%%%%%%%%%%%%%%%%%%%%%%%%%%%%%%%%%%%%%%%%%%%%%%%%%%%%%%%%%
\begin{eqnarray}
	n_{a}^{\rm ss} &=& \frac{4g^2(\Gamma+\gamma)}{ ( 4g^2 + \gamma\Gamma )(\Gamma+\gamma)^2 + 4\gamma\Gamma\Delta_{\rm as}^2 }   
	\cdot \Gamma n_{\rm th},          \\
	n_{b}^{\rm ss} &=& \frac{ 4g^2(\Gamma+\gamma) + \gamma(\Gamma+\gamma)^2 + 4\gamma\Delta_{\rm as}^2 }
	{ ( 4g^2 + \gamma\Gamma )(\Gamma+\gamma)^2 + 4\gamma\Gamma\Delta_{\rm as}^2  }
	\cdot \Gamma n_{\rm th}, 
\end{eqnarray}
%%%%%%%%%%%%%%%%%%%%%%%%%%%%%%%%%%%%%%%%%%%%%%%%%%%%%%%%%%%%%%
where $\Omega = \sqrt{ \sqrt{ \left( 8g^2+2\Delta_{\rm as}^2 - \frac{(\Gamma-\gamma)^2}{2} \right)^2 
		+ 4(\Gamma-\gamma)^2\Delta_{\rm as}^2 } + \left( 8g^2+2\Delta_{\rm as}^2 - \frac{(\Gamma-\gamma)^2}{2} \right) }$
and $n_{a,0}$ ($n_{b,0}$) denotes the initial photon (phonon) number.
We consider the strong coupling regime ($g\gg\Gamma,\gamma$) and assume that the wave-number-induced frequency shift
of anti-Stokes mode is within the linewidth of the acoustic mode ($\Delta_{\rm as}<\Gamma$). In addition, for the backward
Brillouin scattering in a typical optical waveguide, we have $\Gamma\gg\gamma$ and $\Delta_{\rm as}\gg\Delta_{\rm ac}$
since $\upsilon_{\rm o}\gg\upsilon_{\rm as}$. Furthermore, at a low temperature of several kelvins, the thermal phonon
occupation is very low because of the GHz acoustic frequency range.  Under these conditions, the mean photon number and 
mean phonon number can be approximated as
%%%%%%%%%%%%%%%%%%%%%%%%%%%%%%%%%%%%%%%%%%%%%%%%%%%%%%%%%%%%%%
\begin{eqnarray}\label{Approximated mean photon and phonon number}
	n_a(t) &\approx& \frac{n_{a,0}}{2} \left( 1 + \cos(2gt) \right) e^{-\frac{\gamma+\Gamma}{2}t}	
	+ \frac{n_{b,0}}{2} \left( 1 - \cos(2gt) \right) e^{-\frac{\gamma+\Gamma}{2}t}
	+ \left( 1 - e^{-\frac{\gamma+\Gamma}{2}t} \right) n_{\rm th}, \nonumber \\
	n_b(t) &\approx& \frac{n_{a,0}}{2} \left( 1 - \cos(2gt) \right) e^{-\frac{\gamma+\Gamma}{2}t}
	+ \frac{n_{b,0}}{2} \left( 1 + \cos(2gt) \right) e^{-\frac{\gamma+\Gamma}{2}t}
	+ \left( 1 - e^{-\frac{\gamma+\Gamma}{2}t} \right) n_{\rm th}.
\end{eqnarray}
%%%%%%%%%%%%%%%%%%%%%%%%%%%%%%%%%%%%%%%%%%%%%%%%%%%%%%%%%%%%%%

\section{Quantum memory of squeezed states}\label{S2}
In this section, we will discuss the probability of quantum memory for squeezed states, including the squeezed vacuum states,
squeezed thermal states, and squeezed coherent states, via the above quantum transduction.

\subsection{Memory of squeezed vacuum states}
We consider that the initial state of the signal light $a_{\rm sg}$ is prepared to an squeezed vacuum state
$S|0\rangle$, where $S(r)$ is the unitary phase-free squeezed operator with a squeezing degree $r$ which can 
be expressed as $S(r)={\rm exp}\left[ \frac{r}{2}( a_{\rm sg}^2  - (a_{\rm sg}^{\dagger})^2 ) \right]$~\cite{adesso2008,weedbrook2012}. 
For simplification, we assume that the initial state of acoustic phonons $b$ is the ground state. Here, 
$a_{\rm sg}(k)$ and $b(k)$ correspond to the annihilation operators of the signal photons and acoustic phonons
with the wave number $k$, respectively. We define the quadrature operators of signal photons and acoustic phonons 
as follows
%%%%%%%%%%%%%%%%%%%%%%%%%%%%%%%%%%%%%%%%%%%%%%%%%%%%%%%%%%%%%%
\begin{eqnarray}
	X_{\rm sg} &=& \frac{ a_{\rm sg} + a_{\rm sg}^{\dagger} }{\sqrt{2}}, \quad 
	P_{\rm sg} = \frac{ a_{\rm sg} - a_{\rm sg}^{\dagger} }{i\sqrt{2}}, \nonumber\\
	X_{\rm b} &=& \frac{ b^{\dagger}e^{i\beta_b} + b e^{-i\beta_b} }{\sqrt{2}}, \quad
	P_{\rm b} =  \frac{ b^{\dagger}e^{i(\beta_b+\pi/2)} + b e^{-i(\beta_b+\pi/2)} }{\sqrt{2}},
\end{eqnarray}
%%%%%%%%%%%%%%%%%%%%%%%%%%%%%%%%%%%%%%%%%%%%%%%%%%%%%%%%%%%%%%
where $\beta_b$ corresponds to the phase of the acoustic quadrature operators. As the states of photons and
phonons belong to the family of Gaussian states, we can use two $2\times 2$ symmetric covariance matrices 
$\mathcal{V}_{\rm as}$ and $\mathcal{V}_{\rm b}$~\cite{weedbrook2012} to characterize their states as follows
%%%%%%%%%%%%%%%%%%%%%%%%%%%%%%%%%%%%%%%%%%%%%%%%%%%%%%%%%%%%%%
\begin{equation}
	\mathcal{V}_{\rm as} = \left[ 
	\begin{array}{cc}
		\mathcal{V}_{{\rm as},11} & \mathcal{V}_{{\rm as},12} \\
		\mathcal{V}_{{\rm as},12} & \mathcal{V}_{{\rm as},22}
	\end{array}
	\right], \quad	
	\mathcal{V}_{\rm b} = \left[ 
	\begin{array}{cc}
		\mathcal{V}_{{\rm b},11} & \mathcal{V}_{{\rm b},12} \\
		\mathcal{V}_{{\rm b},12} & \mathcal{V}_{{\rm b},22}
	\end{array}
	\right],
\end{equation}
%%%%%%%%%%%%%%%%%%%%%%%%%%%%%%%%%%%%%%%%%%%%%%%%%%%%%%%%%%%%%%
with elements
\begin{equation}
	\mathcal{V}_{ij} = \frac{ \langle\phi_i(t) \phi_j(t)\rangle + \langle\phi_j(t) \phi_i(t)\rangle }{2} - \langle \phi_i(t)\phi_j(t) \rangle,
\end{equation}
where the indexes $i$ and $j$ go over the vector $\phi^T(t)=[ X(t), P(t) ]$. With properties of the squeezed vacuum state,
covariance matrix $\mathcal{V}_{\rm as}$ of signal photons at the initial time can be expressed as
%%%%%%%%%%%%%%%%%%%%%%%%%%%%%%%%%%%%%%%%%%%%%%%%%%%%%%%%%%%%%%
\begin{equation}
	\mathcal{V}_{\rm sg}(t=0) = \left[
	\begin{array}{cc}
		\frac{1}{2} e^{-2r} & 0 \\
		0 & \frac{1}{2} e^{2r}
	\end{array}
	\right].
\end{equation}
%%%%%%%%%%%%%%%%%%%%%%%%%%%%%%%%%%%%%%%%%%%%%%%%%%%%%%%%%%%%%%
In the writing process, the dynamics of the effective anti-Stokes process can be given by
\begin{eqnarray}\label{Motion equation during writing process}
	\frac{d a_{\rm as}}{dt} &=& \left(-\frac{\gamma}{2} + i\Delta_{\rm as} \right) a_{\rm as} - ig_1 b + \sqrt{\gamma}\xi_{\rm as}, \nonumber\\
	\frac{d b}{dt} &=& \left( -\frac{\Gamma}{2} + i\Delta_{ac} \right) b - ig_1 a_{\rm as} + \sqrt{\Gamma}\xi_{ac},
\end{eqnarray}
where $g_1$ denotes the effective optoacoustic coupling strength during the writing process. By solving Eqs.~(\ref{Motion equation during writing process}),
these three independent elements of covariance matrix $\mathcal{V}_{\rm b}$ at time $t$ can be given by
\begin{eqnarray}
	\mathcal{V}_{{\rm b},11}(t) &=& \frac{1}{2} - \frac{1}{2} e^{-2i\beta_b}\mu_1^2\left( e^{\omega_{+}t} - e^{\omega_{-}t} \right)^2 
	\cosh(r)\sinh(r) \nonumber\\
	&& -\frac{1}{2} e^{-2i\beta_b} \left( \mu_1^* \right)^2 \left( e^{\omega_{+}^{*}t} - e^{\omega_{-}^{*}t} \right)^2 \cosh(r)\sinh(r) 
	+ |\mu_1|^2 \left| e^{\omega_{+}t} - e^{\omega_{-}t} \right|^2 	+ \sinh^2(r) \nonumber\\
	&& + \Gamma n_{\rm th} \left[ \frac{|\mu_3|^2}{\alpha_1} \left( e^{\alpha_1 t} - 1 \right)
	- \frac{\mu_3\mu_2^*}{\alpha_2} \left( e^{\alpha_2 t} - 1 \right)
	- \frac{\mu_2\mu_3^*}{\alpha_3} \left( e^{\alpha_3 t} - 1 \right) 
	+ \frac{|\mu_2|^2}{\alpha_4}    \left( e^{\alpha_4 t} - 1 \right)  \right], \nonumber\\
	\mathcal{V}_{{\rm b},22}(t) &=& \frac{1}{2} - \frac{1}{2} e^{-i(2\beta_b+\pi)} \mu_1^2\left( e^{\omega_{+}t} - e^{\omega_{-}t} \right)^2
	\cosh(r)\sinh(r) \nonumber\\   
	&& -\frac{1}{2} e^{-i(2\beta_b+\pi)} \left( \mu_1^* \right)^2 \left( e^{\omega_{+}^{*}t} - e^{\omega_{-}^{*}t} \right)^2 \cosh(r)\sinh(r)  
	+ |\mu_1|^2 \left| e^{\omega_{+}t} - e^{\omega_{-}t} \right|^2 	+ \sinh^2(r) \nonumber\\   
	&& + \Gamma n_{\rm th} \left[ \frac{|\mu_3|^2}{\alpha_1} \left( e^{\alpha_1 t} - 1 \right)
	- \frac{\mu_3\mu_2^*}{\alpha_2} \left( e^{\alpha_2 t} - 1 \right)
	- \frac{\mu_2\mu_3^*}{\alpha_3} \left( e^{\alpha_3 t} - 1 \right) 
	+ \frac{|\mu_2|^2}{\alpha_4}    \left( e^{\alpha_4 t} - 1 \right)  \right],  \nonumber\\
	\mathcal{V}_{{\rm b},12}(t) &=& - \frac{1}{2} e^{-i(2\beta_b+\pi/2)} \mu_1^2 \left( e^{\omega_{+}t} - e^{\omega_{-}t} \right)^2
	\cosh(r)\sinh(r) \nonumber\\
	&& - \frac{1}{2} e^{i(2\beta_b+\pi/2)} \left( \mu_1^* \right)^2 \left( e^{\omega_{+}^{*}t} - e^{\omega_{-}^{*}t} \right)^2 \cosh(r)\sinh(r),
\end{eqnarray}
where
\begin{eqnarray}\label{Coefficients11}
	\omega_{+} &=& -\frac{\gamma+\Gamma}{4} + i\frac{\Delta_{\rm as}+\Delta_{\rm ac}}{2}
	-i \frac{ \sqrt{ 16g_1^2 - \left[ (\Gamma - \gamma) + 2i(\Delta_{\rm as}-\Delta_{\rm ac}) \right]^2 } }{4}, \nonumber\\
	\omega_{-} &=& -\frac{\gamma+\Gamma}{4} + i\frac{\Delta_{\rm as}+\Delta_{\rm ac}}{2}
	+i \frac{ \sqrt{ 16g_1^2 - \left[ (\Gamma - \gamma) + 2i(\Delta_{\rm as}-\Delta_{\rm ac}) \right]^2 } }{4}, \nonumber\\ 
	\tau_{+}   &=& \frac{-2(\Delta_{\rm as}-\Delta_{\rm ac})+i(\Gamma-\gamma)}{4g_1} 
	+  \frac{ \sqrt{ 16g_1^2 - \left[ (\Gamma - \gamma) + 2i(\Delta_{\rm as}-\Delta_{\rm ac}) \right]^2 } }{4g_1}, \nonumber\\
	\tau_{-}   &=& \frac{-2(\Delta_{\rm as}-\Delta_{\rm ac})+i(\Gamma-\gamma)}{4g_1}
	-  \frac{ \sqrt{ 16g_1^2 - \left[ (\Gamma - \gamma) + 2i(\Delta_{\rm as}-\Delta_{\rm ac}) \right]^2 } }{4g_1},       
\end{eqnarray}
and
\begin{eqnarray}\label{Coefficients12}
	\mu_1 &=& \frac{1}{\tau_{+} - \tau_{-}}, \quad \mu_2 = \frac{\tau_{+}}{\tau_{+} - \tau_{-}}, \quad \mu_3 = \frac{\tau_{-}}{\tau_{+} - \tau_{-}}, \nonumber\\
	\alpha_1 &=& \omega_{+} + \omega_{+}^{*}, \quad \alpha_2 = \omega_{+} + \omega_{-}^{*}, \nonumber\\
	\alpha_3 &=& \omega_{-} + \omega_{+}^{*}, \quad \alpha_4 = \omega_{-} + \omega_{-}^{*}.
\end{eqnarray}
The displacement quadrature $X_{\rm b}$ can be squeezed at time $t$ with phase $\beta_b=\pi/2$, where the corresponding
variance can be expressed as $\Delta^2 X_{\rm b}(t) = \mathcal{V}_{{\rm b},11}(t)$, which can be approximately expressed as
\begin{equation}
	\Delta^2 X_{\rm b}(t) \approx \frac{1}{2} - \frac{1-e^{-2r}}{2} e^{-\frac{\Gamma}{2}t} \sin^2(g_1 t) + \frac{\Gamma n_{\rm th}}{2} t.
\end{equation}
We can see that the variance of displace quadrature $X_{\rm b}$ experiences a Rabi oscillation with frequency $\sim2g_1$, and thus its minimum value
can be obtained at the end of the first half Rabi oscillation, i.e., $\tau_1\approx \pi/(2g_1)$, which can be given by
\begin{equation}
	\Delta^2 X_{\rm b}^{\rm min} \approx \frac{1}{2} e^{-2r} + \frac{\pi\Gamma}{4g_1}\left( n_{\rm th} + \frac{1-e^{-2r}}{2} \right).
\end{equation}
Now we calculate the fidelity between signal photons and acoustic phonons to quantify the state transfer during the writing process.
The fidelity between two single-mode Gaussian states $\rho_{1,2}$ can be defined as~\cite{adesso2008,chiribella2014} 
\begin{equation}
	\mathcal{F}\left[ \rho_1, \rho_2 \right] = \frac{1}{ { \sqrt{\Delta+\Lambda} - \sqrt{\Lambda} }}
	\exp\left[ -\frac{1}{4}\delta_u^T ( \mathcal{V}_1 + \mathcal{V}_2 )^{-1} \delta_u \right],
\end{equation}
where $\mathcal{V}_{1,2}$ correspond to the covariance matrices of $\rho_{1,2}$ and mean value $\delta_u^T=[\langle X_1-X_2\rangle, \langle P_1-P_2\rangle]$. 
As the mean values of $\langle X_{\rm as} - X_{\rm b} \rangle=\langle P_{\rm as} - P_{\rm b}  \rangle=0$, the fidelity between
the initial state of signal photons and state of acoustic phonons at time $t$ during the writing process can be given by
\begin{eqnarray}
	\mathcal{F}\left[ \mathcal{V}_{\rm as}(0), \mathcal{V}_{\rm b}(t) \right] = \frac{1}{ { \sqrt{\Delta+\Lambda} - \sqrt{\Lambda} }},
\end{eqnarray}
with
\begin{eqnarray}
	\Delta  &=& \det\left[ \mathcal{V}_{\rm as}(0) + \mathcal{V}_{\rm b}(t) \right], \nonumber\\
	\Lambda &=& 4\left[ \det[\mathcal{V}_{\rm as}(0)] - \frac{1}{4} \right] \left[ \det[\mathcal{V}_{\rm b}(t)] - \frac{1}{4} \right].
\end{eqnarray}
According to the above discussion, we can see that it is capable of transferring the squeezed vacuum state from the signal photons to
the acoustic phonons via a Brillouin anti-Stokes process, and a high fidelity is possible to be achieved at the optimal time 
$\tau_1\approx\pi/(2g_1)$.

Now we analyze the performance of the storage of this squeezed state on acoustic phonons during the storage process. We switch off the 
first pump at optimal time $\tau_1$, and thereby the acoustic phonons are driven by the thermal noise during the storage process, where
the dynamics can be given by
\begin{eqnarray}\label{Dynamics during the storage process}
	\frac{d b}{dt} = (-\frac{\Gamma}{2}+i\Delta_{\rm ac}) b + \sqrt{\Gamma}\xi_{\rm ac}.
\end{eqnarray}
By solve Eq.~(\ref{Dynamics during the storage process}), the elements of acoustic covariance matrix $\bar{\mathcal{V}}_{\rm b}$ can be written 
analytically as follows
\begin{eqnarray}
	\bar{\mathcal{V}}_{{\rm b},11}(t) &=& \frac{1}{2} - \frac{1}{2} e^{-2i\beta_b} e^{2(-\frac{\Gamma}{2}+i\Delta_{\rm ac})t}
	\mu_1^2 \left( e^{\omega_{+}\tau_1} - e^{\omega_{-}\tau_1} \right)^2\cosh(r)\sinh(r) \nonumber\\
	&& - \frac{1}{2} e^{2i\beta_b} e^{2(-\frac{\Gamma}{2}-i\Delta_{\rm ac})t} (\mu_1^*)^2
	\left( e^{\omega_{+}^{*}\tau_1} - e^{\omega_{-}^{*}\tau_1} \right)^2\cosh(r)\sinh(r) \nonumber\\
	&& + n_{\rm th}\left( 1 - e^{-\Gamma t} \right) 
	+ e^{-\Gamma t} |\mu_1|^2 \left| e^{\omega_{+}\tau_1} - e^{\omega_{-}\tau_1} \right|^2 \sinh^2(r) \nonumber\\
	&& + \Gamma n_{\rm th} e^{-\Gamma t} \left[ \frac{|\mu_3|^2}{\alpha_1} \left( e^{\alpha_1\tau_1} - 1 \right)
	- \frac{\mu_3\mu_2^{*}}{\alpha_2} \left( e^{\alpha_2\tau_1} - 1 \right)
	- \frac{\mu_2\mu_3^{*}}{\alpha_3} \left( e^{\alpha_3\tau_1} - 1 \right)
	+ \frac{|\mu_2|^2}{\alpha_4} \left( e^{\alpha_4\tau_1} - 1 \right) \right], \nonumber\\
	\bar{\mathcal{V}}_{{\rm b},22}(t) &=& \frac{1}{2} - \frac{1}{2} e^{-2i(\beta_b+\pi/2)} e^{2(-\frac{\Gamma}{2}+i\Delta_{\rm ac})t}
	\mu_1^2 \left( e^{\omega_{+}\tau_1} - e^{\omega_{-}\tau_1} \right)^2\cosh(r)\sinh(r) \nonumber\\
	&& - \frac{1}{2} e^{2i(\beta_b+\pi/2)} e^{2(-\frac{\Gamma}{2}-i\Delta_{\rm ac})t} (\mu_1^*)^2
	\left( e^{\omega_{+}^{*}\tau_1} - e^{\omega_{-}^{*}\tau_1} \right)^2\cosh(r)\sinh(r) \nonumber\\
	&& + n_{\rm th}\left( 1 - e^{-\Gamma t} \right) 
	+ e^{-\Gamma t} |\mu_1|^2 \left| e^{\omega_{+}\tau_1} - e^{\omega_{-}\tau_1} \right|^2 \sinh^2(r) \nonumber\\
	&& + \Gamma n_{\rm th} e^{-\Gamma t} \left[ \frac{|\mu_3|^2}{\alpha_1} \left( e^{\alpha_1\tau_1} - 1 \right)
	- \frac{\mu_3\mu_2^{*}}{\alpha_2} \left( e^{\alpha_2\tau_1} - 1 \right)
	- \frac{\mu_2\mu_3^{*}}{\alpha_3} \left( e^{\alpha_3\tau_1} - 1 \right)
	+ \frac{|\mu_2|^2}{\alpha_4} \left( e^{\alpha_4\tau_1} - 1 \right) \right], \nonumber\\
	\bar{\mathcal{V}}_{{\rm b},12}(t) &=& -\frac{1}{2} e^{-i(2\beta_b+\pi/2)} e^{2(-\frac{\Gamma}{2}+i\Delta_{\rm ac})t}
	\mu_1^2 \left( e^{\omega_{+}\tau_1} - e^{\omega_{-}\tau_1} \right)^2\cosh(r)\sinh(r) \nonumber\\
	&& - \frac{1}{2} e^{i(2\beta_b+\pi/2)} e^{2(-\frac{\Gamma}{2}-i\Delta_{\rm ac})t} (\mu_1^*)^2 
	\left( e^{\omega_{+}^{*}\tau_1} - e^{\omega_{-}^{*}\tau_1} \right)^2\cosh(r)\sinh(r).   
\end{eqnarray}
The variance of the squeezed displace quadrature during storage process can be expressed as $\Delta^2 X_{\rm b}(t)=\bar{\mathcal{V}}_{{\rm b},11}(t)$.
The fidelity between initial state of signal photons and state of acoustic phonons at time $t$ during the storage process 
can be expressed as
\begin{eqnarray}
	\bar{\mathcal{F}}[\mathcal{V}_{\rm as}(0), \bar{\mathcal{V}}_{\rm b}(t)]=\frac{1}{ { \sqrt{\Delta+\Lambda} - \sqrt{\Lambda} }},
\end{eqnarray}
with
\begin{eqnarray}
	\Delta  &=& \det\left[ \mathcal{V}_{\rm as}(0) + \bar{\mathcal{V}}_{\rm b}(t) \right], \nonumber\\
	\Lambda &=& 4\left[ \det[\mathcal{V}_{\rm as}(0)] - \frac{1}{4} \right] \left[ \det[\bar{\mathcal{V}}_{\rm b}(t)] - \frac{1}{4} \right].	
\end{eqnarray}

Finally, after a storage interval $\tau_{\rm s}$, we apply a second pump to the waveguide and map the stored quantum state into
an retrieval light $a_{\rm re}$. The dynamics of linearized optoacoustic interaction during the readout process can be given by
\begin{eqnarray}\label{Dynamics of optoacoustic interaction during readout process}
	\frac{d a_{\rm re}}{dt} &=& \left( - \frac{\gamma}{2} + i\Delta_{\rm re} \right) a_{\rm re} - ig_2 b + \sqrt{\gamma}\xi_{\rm re}, \nonumber\\
	\frac{db}{dt} &=& \left( -\frac{\Gamma}{2} + i\Delta_{\rm ac} \right) b - ig_2 a_{\rm re} + \sqrt{\Gamma}\xi_{\rm ac},
\end{eqnarray}
where $g_2$ quantifies the effective optoacoustic coupling strength. Here, we consider the initial vacuum state for retrieval photons. 
We define the quadrature operators of retrieval photons as
$X_{\rm re}=( a_{\rm re}^{\dagger} e^{i\beta_{\rm re}} + a_{\rm re} e^{-i\beta_{\rm re}} )/\sqrt{2}$,
$P_{\rm re}=( a_{\rm re}^{\dagger} e^{i(\beta_{\rm re}+\pi/2)} + a_{\rm re} e^{-i(\beta_{\rm re}+\pi/2) }/(i\sqrt{2})$ with phase $\beta_{\rm re}$.
Solving Eqs.~(\ref{Dynamics of optoacoustic interaction during readout process}),
the covariance matrix $\mathcal{V}_{{\rm re}}$ of retrieval photons at time $t$ can be given by
\begin{eqnarray}
	\mathcal{V}_{{\rm re},11}(t) &=& \frac{1}{2} - \frac{1}{2} e^{-2i\beta_{\rm re}} \tilde{\mu}_1^2
	\left( e^{\tilde{\omega}_{+}t} - e^{\tilde{\omega}_{-}t} \right)^2 e^{2(-\frac{\Gamma}{2}+i\Delta_{\rm ac})\tau_{\rm s}}
	\mu_1^2 \left( e^{\omega_{+}\tau_1} - e^{\omega_{-}\tau_1} \right)^2 \cosh(r)\sinh(r) \nonumber\\
	&& - \frac{1}{2} e^{2i\beta_{\rm re}} \left( \tilde{\mu}_1^{*} \right)^2 
	\left( e^{\tilde{\omega}_{+}^{*}t} - e^{\tilde{\omega}_{-}^{*}t} \right)^2 e^{2(-\frac{\Gamma}{2}-i\Delta_{\rm ac})\tau_{\rm s}} 
	\left( \mu_1^{*} \right)^2 \left( e^{\omega_{+}^{*}\tau_1} - e^{\omega_{-}^{*}\tau_1} \right)^2 \cosh(r)\sinh(r) \nonumber\\
	&& + |\tilde{\mu}_1|^2 \left| e^{\tilde{\omega}_{+}t} - e^{\tilde{\omega}_{-}t} \right|^2 
	\left\langle \tilde{b}^{\dagger}(0) \tilde{b}(0) \right\rangle \nonumber\\
	&& + \Gamma n_{\rm th} |\tilde{\mu}_1|^2 \left[ \frac{1}{\tilde{\alpha}_1} \left( e^{\tilde{\alpha}_1 t} - 1 \right) 
	- \frac{1}{\tilde{\alpha}_2} \left( e^{\tilde{\alpha}_2 t} - 1 \right)
	- \frac{1}{\tilde{\alpha}_3} \left( e^{\tilde{\alpha}_3 t} - 1 \right)
	+ \frac{1}{\tilde{\alpha}_4} \left( e^{\tilde{\alpha}_4 t} - 1 \right) \right], \nonumber\\
	\mathcal{V}_{{\rm re},22}(t) &=& \frac{1}{2} - \frac{1}{2} e^{-2i(\beta_{\rm re}+\pi/2)} \tilde{\mu}_1^2
	\left( e^{\tilde{\omega}_{+}t} - e^{\tilde{\omega}_{-}t} \right)^2 e^{2(-\frac{\Gamma}{2}+i\Delta_{\rm ac})\tau_{\rm s}}	
	\mu_1^2 \left( e^{\omega_{+}\tau_1} - e^{\omega_{-}\tau_1} \right)^2 \cosh(r)\sinh(r) \nonumber\\	  
	&& - \frac{1}{2} e^{2i(\beta_{\rm re}+\pi/2)} \left( \tilde{\mu}_1^{*} \right)^2 
	\left( e^{\tilde{\omega}_{+}^{*}t} - e^{\tilde{\omega}_{-}^{*}t} \right)^2 e^{2(-\frac{\Gamma}{2}-i\Delta_{\rm ac} )\tau_{\rm s}} 
	\left( \mu_1^{*} \right)^2 \left( e^{\omega_{+}^{*}\tau_1} - e^{\omega_{-}^{*}\tau_1} \right)^2 \cosh(r)\sinh(r) \nonumber\\    
	&& + |\tilde{\mu}_1|^2 \left| e^{\tilde{\omega}_{+}t} - e^{\tilde{\omega}_{-}t} \right|^2
	\left\langle \tilde{b}^{\dagger}(0) \tilde{b}(0) \right\rangle \nonumber\\   
	&& + \Gamma n_{\rm th} |\tilde{\mu}_1|^2 \left[ \frac{1}{\tilde{\alpha}_1} \left( e^{\tilde{\alpha}_1 t} - 1 \right) 
	- \frac{1}{\tilde{\alpha}_2} \left( e^{\tilde{\alpha}_2 t} - 1 \right)
	- \frac{1}{\tilde{\alpha}_3} \left( e^{\tilde{\alpha}_3 t} - 1 \right)
	+ \frac{1}{\tilde{\alpha}_4} \left( e^{\tilde{\alpha}_4 t} - 1 \right) \right], \nonumber\\ 
	\mathcal{V}_{{\rm re},12}(t) &=& -\frac{1}{2} e^{-i(2\beta_{\rm re}+\pi/2)} \tilde{\mu}_1^2    
	\left( e^{\tilde{\omega}_{+}t} - e^{\tilde{\omega}_{-}t} \right)^2 e^{2(-\frac{\Gamma}{2}+i\Delta_{\rm ac})\tau_{\rm s}}    
	\mu_1^2  \left( e^{\omega_{+}\tau_1} - e^{\omega_{-}\tau_1} \right)^2 \cosh(r)\sinh(r) \nonumber\\  
	&&	 -\frac{1}{2} e^{i(2\beta_{\rm re}+\pi/2)} \left( \tilde{\mu}_1^{*} \right)^2  
	\left( e^{\tilde{\omega}_{+}^{*}t} - e^{\tilde{\omega}_{-}^{*}t} \right)^2 e^{2(-\frac{\Gamma}{2}-i\Delta_{\rm ac} )\tau_{\rm s}} 
	\left( \mu_1^{*} \right)^2 \left( e^{\omega_{+}^{*}\tau_1} - e^{\omega_{-}^{*}\tau_1} \right)^2 \cosh(r)\sinh(r),                  	            
\end{eqnarray}
with initial acoustic state in readout process
\begin{eqnarray}
	\left\langle \tilde{b}^{\dagger}(0) \tilde{b}(0) \right\rangle &=& n_{\rm th} \left( 1 - e^{-\Gamma\tau_{\rm s}} \right)
	+ e^{-\Gamma\tau_{\rm s}} |\mu_1|^2 \left| e^{\omega_{+}\tau_1} - e^{\omega_{-}\tau_1} \right|^2 \sinh^2(r) \nonumber\\
	&& + \Gamma n_{\rm th} e^{-\Gamma\tau_{\rm s}} \left[ \frac{|\mu_3|^2}{\alpha_1} \left( e^{\alpha_1\tau_1} - 1 \right) 
	- \frac{\mu_3\mu_2^{*}}{\alpha_2} \left( e^{\alpha_2\tau_1} - 1 \right) 
	- \frac{\mu_2\mu_3^{*}}{\alpha_3} \left( e^{\alpha_3\tau_1} - 1 \right)
	+ \frac{|\mu_2|^2}{\alpha_4} \left( e^{\alpha_4\tau_1} - 1 \right) \right]
\end{eqnarray}
and
\begin{eqnarray} \label{Coefficents21}
	\tilde{\omega}_{+} &=& - \frac{\Gamma+\gamma}{4} + i\frac{\Delta_{\rm re}+\Delta_{\rm ac}}{2}
	-i \frac{ \sqrt{ 16g_2^2 - [ (\Gamma-\gamma) + 2i(\Delta_{\rm re}-\Delta_{\rm ac}) ]^2 } }{4}, \nonumber\\
	\tilde{\omega}_{-} &=& - \frac{\Gamma+\gamma}{4} + i\frac{\Delta_{\rm re}+\Delta_{\rm ac}}{2}
	+i \frac{ \sqrt{ 16g_2^2 - [ (\Gamma-\gamma) + 2i(\Delta_{\rm re}-\Delta_{\rm ac}) ]^2 } }{4}, \nonumber\\ 
	\tilde{\tau}_{+}   &=& \frac{ -2(\Delta_{\rm re}-\Delta_{\rm ac}) + i(\Gamma-\gamma) }{4g_2}	   
	+ \frac{ \sqrt{ 16g_2^2 - [ (\Gamma-\gamma) + 2i(\Delta_{\rm re}-\Delta_{\rm ac}) ]^2 } }{4g_2}, \nonumber\\
	\tilde{\tau}_{-}   &=& \frac{ -2(\Delta_{\rm re}-\Delta_{\rm ac}) + i(\Gamma-\gamma) }{4g_2}
	- \frac{ \sqrt{ 16g_2^2 - [ (\Gamma-\gamma) + 2i(\Delta_{\rm re}-\Delta_{\rm ac}) ]^2 } }{4g_2},                                                              
\end{eqnarray}
and
\begin{eqnarray} \label{Coefficents22}
	\tilde{\mu}_1 &=& \frac{1}{\tilde{\tau}_{+} - \tilde{\tau}_{-}}, \quad 
	\tilde{\mu}_2 = \frac{\tilde{\tau}_{+}}{\tilde{\tau}_{+} - \tilde{\tau}_{-}}, \quad
	\tilde{\mu}_3 = \frac{\tilde{\tau}_{-}}{\tilde{\tau}_{+} - \tilde{\tau}_{-}}, \nonumber\\
	\tilde{\alpha}_1 &=&  \tilde{\omega}_{+} + \tilde{\omega}_{+}^{*}, \quad
	\tilde{\alpha}_2 = \tilde{\omega}_{+} + \tilde{\omega}_{-}^{*},	\nonumber\\
	\tilde{\alpha}_3 &=&  \tilde{\omega}_{-} + \tilde{\omega}_{+}^{*}, \quad   
	\tilde{\alpha}_4 = \tilde{\omega}_{-} + \tilde{\omega}_{-}^{*}.               
\end{eqnarray}
We choose displacement quadrature phase of retrieval photons as $\beta_{\rm re}=0$, thus the variance $\Delta^2 X_{\rm re}$ 
can be given by
\begin{eqnarray}
	\Delta^2 X_{\rm re}(t) &=& \mathcal{V}_{{\rm re},11}(t) \nonumber\\
	&=& \frac{1}{2} - \frac{1}{2} \tilde{\mu}_1^2
	\left( e^{\tilde{\omega}_{+}t} - e^{\tilde{\omega}_{-}t} \right)^2 e^{2(-\frac{\Gamma}{2}+i\Delta_{\rm ac})\tau_{\rm s}}
	\mu_1^2 \left( e^{\omega_{+}\tau_1} - e^{\omega_{-}\tau_1} \right)^2 \cosh(r)\sinh(r) \nonumber\\
	&& - \frac{1}{2} \left( \tilde{\mu}_1^{*} \right)^2
	\left( e^{\tilde{\omega}_{+}^{*}t} - e^{\tilde{\omega}_{-}^{*}t} \right)^2 e^{2(-\frac{\Gamma}{2}-i\Delta_{\rm ac})\tau_{\rm s}}
	\left( \mu_1^{*} \right)^2 \left( e^{\omega_{+}^{*}\tau_1} - e^{\omega_{-}^{*}\tau_1} \right)^2 \cosh(r)\sinh(r) \nonumber\\
	&& + |\tilde{\mu}_1|^2 \left| e^{\tilde{\omega}_{+}t} - e^{\tilde{\omega}_{-}t} \right|^2
	\left\langle \tilde{b}^{\dagger}(0) \tilde{b}(0) \right\rangle \nonumber\\
	&& + \Gamma n_{\rm th} |\tilde{\mu}_1|^2 \left[ \frac{1}{\tilde{\alpha}_1} \left( e^{\tilde{\alpha}_1 t} - 1 \right)
	- \frac{1}{\tilde{\alpha}_2} \left( e^{\tilde{\alpha}_2 t} - 1 \right)
	- \frac{1}{\tilde{\alpha}_3} \left( e^{\tilde{\alpha}_3 t} - 1 \right)
	+ \frac{1}{\tilde{\alpha}_4} \left( e^{\tilde{\alpha}_4 t} - 1 \right) \right].
\end{eqnarray}
Considering a short storage time, i.e., $\tau_{\rm s}\ll1/\Gamma$, the variance $\Delta^2 X_{\rm re}$ can be approximately expressed
\begin{eqnarray}
	\Delta^2 X_{\rm re} &\approx& \frac{1}{2} - \frac{1-e^{-2r}}{2}e^{-\frac{\Gamma}{2}t}\sin^2(g_2t)
	+ \Gamma n_{\rm th} \left[ \frac{t}{2} + \frac{\pi}{4g_1} e^{-\frac{\Gamma}{2}t} \sin^2(g_2t) \right].
\end{eqnarray}
The property of the Rabi oscillation indicates that the minimum value of $\Delta^2 X_{\rm re}(t)$ can be achieved at
time $\tau_2\approx\pi/(2g_2)$ and be simplified as follows
\begin{eqnarray}
	\Delta^2 X_{\rm re}^{\rm min} \approx \frac{1}{2} e^{-2r} + \frac{\pi}{4} 
	\left[ \frac{\Gamma n_{\rm th}}{g_1} + \frac{\Gamma n_{\rm th}}{g_2} 
	+ \frac{\Gamma}{2g_2} (1-e^{-2r}) \right].
\end{eqnarray}
The fidelity between the initial state $\mathcal{V}_{\rm sg}(0)$ of signal photons and the state $\mathcal{V}_{\rm re}(t)$ 
of retrieval photons during the readout process can be given by
\begin{eqnarray}
	\tilde{\mathcal{F}}\left[ \mathcal{V}_{\rm sg}(0), \mathcal{V}_{\rm re}(t) \right]
	= \frac{1}{  { \sqrt{ \Delta - \Lambda } - \sqrt{\Lambda} } },
\end{eqnarray}
where
\begin{eqnarray}
	\Delta &=& \det\left[ \mathcal{V}_{\rm sg}(0) + \mathcal{V}_{\rm re}(t) \right], \nonumber\\
	\Lambda &=& 4 \left[ \det[ \mathcal{V}_{\rm sg}(0)] - \frac{1}{4} \right] \left[ \det[ \mathcal{V}_{\rm re}(t)] - \frac{1}{4} \right].
\end{eqnarray}

\subsection{Memory of squeezed thermal states}
In this subsection, we will explore the quantum memory of squeezed thermal states in Brillouin-active waveguides. 
We assume that the initial state of the signal light is prepared to a squeezed thermal state 
$\rho_{\rm st}=S^{\dagger}(r)\rho_{\rm th}S(r)$, which is defined by the action of a squeezing operator
$S(r)={\rm exp}\left[ \frac{r}{2}( a_{\rm sg}^2  - (a_{\rm sg}^{\dagger})^2 ) \right]$ on a thermal state~\cite{adesso2008} 
\begin{eqnarray}
	\rho_{\rm th} = \left[ 1 - \exp\left( - \frac{\hbar\omega_{\rm sg}}{k_{\rm B}T_{0}} \right) \right]
	\sum_{n=0}^{\infty} \exp \left( - \frac{n \hbar\omega_{\rm sg}}{k_{\rm B}T_{0}} \right)
	|n\rangle \langle n|,
\end{eqnarray}
where $\omega_{\rm sg}$ denotes the optical frequency and $T_0$ indicates the temperature. The corresponding 
thermal photon occupation can be expressed as
\begin{eqnarray}
	\bar{n}_{\rm th} &=& \left[ \exp\left( \frac{\hbar\omega_{\rm sg}}{k_{\rm B}T_{0}} \right) - 1 \right]^{-1} \nonumber\\
	&=& \frac{1}{2} \left( \frac{1}{2u} - 1 \right).
\end{eqnarray}
When $u=1$, the squeezed thermal state will reduce to the squeezed vacuum state $S(r)|0\rangle$.
Thus the covariance matrix of the signal photons at initial time $t=0$ can be expressed as
\begin{eqnarray}
	\mathcal{V}_{\rm sg}(0) = \left[
	\begin{array}{cc}
		\frac{1}{2u} e^{-2r} & 0 \\
		0 & \frac{1}{2u} e^{2r}
	\end{array}
	\right].
\end{eqnarray}

After transferred the squeezed thermal state from the signal photons to the acoustic phonons in the writing process, 
the elements of covariance matrix $\mathcal{V}_{\rm b}(t)$ for the acoustic phonons can be given by
\begin{eqnarray}
	\mathcal{V}_{{\rm b},11}(t) &=& \frac{1}{2} - \frac{1}{2} e^{-2i\beta_b} \mu_1^2
	\left( e^{\omega_{+}t} - e^{\omega_{-}t} \right)^2 (2\bar{n}_{\rm th}+1) \cosh(r)\sinh(r) \nonumber\\
	&& - \frac{1}{2} e^{2i\beta_b} (\mu_1^{*})^2 \left( e^{\omega_{+}^{*}t} - e^{\omega_{-}^{*}t} \right)^2
	(2\bar{n}_{\rm th}+1) \cosh(r)\sinh(r) \nonumber\\
	&& + |\mu_1|^2 \left| e^{\omega_{+}t} - e^{\omega_{-}t} \right|^2
	\left[ \bar{n}_{\rm th} \left( \cosh^2(r) + \sinh^2(r) \right) + \sinh^2(r) \right] \nonumber\\
	&& + \Gamma n_{\rm th} \left[ \frac{|\mu_3|^2}{\alpha_1} \left( e^{\alpha_1 t} - 1 \right) 
	- \frac{\mu_3\mu_2^{*}}{\alpha_2} ( e^{\alpha_2 t} - 1 ) 
	- \frac{\mu_2\mu_3^{*}}{\alpha_3} ( e^{\alpha_3 t} - 1 ) \right], \nonumber\\
	\mathcal{V}_{{\rm b},22}(t) &=& \frac{1}{2} - \frac{1}{2} e^{-i(2\beta_b+\pi)} \mu_1^2
	\left( e^{\omega_{+}t} - e^{\omega_{-}t} \right)^2 (2\bar{n}_{\rm th}+1) \cosh(r)\sinh(r) \nonumber\\
	&& - \frac{1}{2} e^{i(2\beta_b+\pi)} (\mu_1^{*})^2 \left( e^{\omega_{+}^{*}t} - e^{\omega_{-}^{*}t} \right)^2
	(2\bar{n}_{\rm th}+1) \cosh(r)\sinh(r) \nonumber\\
	&& + |\mu_1|^2 \left| e^{\omega_{+}t} - e^{\omega_{-}t} \right|^2
	\left[ \bar{n}_{\rm th} \left( \cosh^2(r) + \sinh^2(r) \right) + \sinh^2(r) \right] \nonumber\\
	&& + \Gamma n_{\rm th} \left[ \frac{|\mu_3|^2}{\alpha_1} \left( e^{\alpha_1 t} - 1 \right) 
	- \frac{\mu_3\mu_2^{*}}{\alpha_2} ( e^{\alpha_2 t} - 1 ) 
	- \frac{\mu_2\mu_3^{*}}{\alpha_3} ( e^{\alpha_3 t} - 1 ) \right], \nonumber\\	
	\mathcal{V}_{{\rm b},12}(t) &=& - \frac{1}{2} e^{-i(2\beta_b+\pi/2)} \mu_1^2
	\left( e^{\omega_{+}t} - e^{\omega_{-}t} \right)^2 (2\bar{n}_{\rm th}+1) \cosh(r)\sinh(r) \nonumber\\  
	&& - \frac{1}{2} e^{i(2\beta_b+\pi/2)} (\mu_1^{*})^2 \left( e^{\omega_{+}^{*}t} - e^{\omega_{-}^{*}t} \right)^2
	(2\bar{n}_{\rm th}+1) \cosh(r)\sinh(r),       
\end{eqnarray}
where coefficients $\omega_{\pm}$, $\tau_{\pm}$, $\mu_{i}$, and $\alpha_{j}$ are illustrated in Eqs.~(\ref{Coefficients11})
and (\ref{Coefficients12}). The variance of the acoustic displacement quadrature is $\Delta^2 X_{\rm b}(t)=\mathcal{V}_{{\rm b},11}(t)$. 
Thus the fidelity between the initial state $\mathcal{V}_{\rm sg}(0)$ of signal photons
and the state $\mathcal{V}_{\rm b}(t)$ of acoustic phonons during the writing process can be given by
\begin{eqnarray}
	\mathcal{F} \left[ \mathcal{V}_{\rm sg}(0), \mathcal{V}_{\rm b}(t) \right]
	= \frac{1}{  { \sqrt{\Delta+\Lambda} - \sqrt{\Lambda} } },
\end{eqnarray}
with
\begin{eqnarray}
	\Delta &=& \det \left[ \mathcal{V}_{\rm sg}(0) + \mathcal{V}_{\rm b}(t) \right], \nonumber\\
	\Lambda &=& 4\left[ \det[\mathcal{V}_{\rm as}(0)] - \frac{1}{4} \right] \left[ \det[\mathcal{V}_{\rm b}(t)] - \frac{1}{4} \right].
\end{eqnarray}
In the strong coupling regime, the state swapping between the signal photons and acoustic phonons experiences 
a Rabi oscillation with frequency $\sim 2g_1$, thereby the state can be transferred to the acoustic phonons with
a high fidelity at the end of the first half Rabi oscillation, i.e., $\tau_1\approx \pi/(2g_1)$.

In the storage process, the symmetric covariance matrix of the acoustic phonons can be given by
\begin{eqnarray}
	\bar{\mathcal{V}}_{{\rm b},11}(t) &=& \frac{1}{2} - \frac{1}{2} e^{-2i\beta_{\rm ac}}
	\mu_1^2 \left( e^{\omega_{+}\tau_1} - e^{\omega_{-}\tau_1} \right)^2
	(2\bar{n}_{\rm th}+1) e^{\left( -\frac{\Gamma}{2} + i\Delta_{\rm ac} \right)t} \cosh(r)\sinh(r) \nonumber\\
	&& - \frac{1}{2} e^{2i\beta_b} \left(\mu_1^{*}\right)^2
	\left( e^{\omega_{+}^{*}\tau_1} - e^{\omega_{-}^{*}\tau_1} \right)^2
	(2\bar{n}_{\rm th}+1) e^{\left( -\frac{\Gamma}{2} - i\Delta_{\rm ac} \right)t} \cosh(r)\sinh(r) \nonumber\\
	&& + n_{\rm th} \left( 1 - e^{-\Gamma t} \right)
	+ e^{-\Gamma t} |\mu_1|^2 \left| e^{\omega_{+}\tau_1} - e^{\omega_{-}\tau_1} \right|^2
	\left[ \bar{n}_{\rm th} \left( \cosh^2(r) + \sinh^2(r) \right) + \sinh^2(r) \right]  \nonumber \\
	&& + \Gamma \bar{n}_{\rm th} e^{-\Gamma t} \left[ 
	\frac{|\mu_3|^2}{\alpha_1} \left( e^{\alpha_1\tau_1} - 1 \right) 
	- \frac{\mu_3\mu_2^{*}}{\alpha_2} \left( e^{\alpha_3\tau_1} - 1 \right) 
	- \frac{\mu_2\mu_3^{*}}{\alpha_3} \left( e^{\alpha_3\tau_1} - 1 \right)
	+ \frac{|\mu_2|^2}{\alpha_4} \left( e^{\alpha_4\tau_1} - 1 \right) \right], \nonumber\\
	\bar{\mathcal{V}}_{{\rm b},22}(t) &=& \frac{1}{2} - \frac{1}{2} e^{-i(2\beta_b+\pi)}	 
	\mu_1^2 \left( e^{\omega_{+}\tau_1} - e^{\omega_{-}\tau_1} \right)^2  
	(2\bar{n}_{\rm th}+1) e^{\left( -\frac{\Gamma}{2} + i\Delta_{\rm ac} \right)t} \cosh(r)\sinh(r) \nonumber\\  
	&& - \frac{1}{2} e^{i(2\beta_b+\pi)} \left(\mu_1^{*}\right)^2   
	\left( e^{\omega_{+}^{*}\tau_1} - e^{\omega_{-}^{*}\tau_1} \right)^2  
	(2\bar{n}_{\rm th}+1) e^{\left( -\frac{\Gamma}{2} - i\Delta_{\rm ac} \right)t} \cosh(r)\sinh(r) \nonumber\\
	&& + n_{\rm th} \left( 1 - e^{-\Gamma t} \right)
	+ e^{-\Gamma t} |\mu_1|^2 \left| e^{\omega_{+}\tau_1} - e^{\omega_{-}\tau_1} \right|^2   
	\left[ \bar{n}_{\rm th} \left( \cosh^2(r) + \sinh^2(r) \right) + \sinh^2(r) \right]  \nonumber \\ 
	&&  + \Gamma \bar{n}_{\rm th} e^{-\Gamma t} \left[
	\frac{|\mu_3|^2}{\alpha_1} \left( e^{\alpha_1\tau_1} - 1 \right)
	- \frac{\mu_3\mu_2^{*}}{\alpha_2} \left( e^{\alpha_3\tau_1} - 1 \right)
	- \frac{\mu_2\mu_3^{*}}{\alpha_3} \left( e^{\alpha_3\tau_1} - 1 \right)
	+ \frac{|\mu_2|^2}{\alpha_4} \left( e^{\alpha_4\tau_1} - 1 \right) \right], \nonumber\\ 
	\bar{\mathcal{V}}_{{\rm b},12}(t) &=& -\frac{1}{2} e^{-i(2\beta_b+\pi/2)} 	
	\mu_1^2 \left( e^{\omega_{+}\tau_1} - e^{\omega_{-}\tau_1} \right)^2 
	(2\bar{n}_{\rm th}+1) e^{\left( -\frac{\Gamma}{2} + i\Delta_{\rm ac} \right)t} \cosh(r)\sinh(r) \nonumber\\ 
	&& -\frac{1}{2} e^{i(2\beta_b+\pi/2)} \left(\mu_1^{*}\right)^2 
	\left( e^{\omega_{+}^{*}\tau_1} - e^{\omega_{-}^{*}\tau_1} \right)^2   
	(2\bar{n}_{\rm th}+1) e^{\left( -\frac{\Gamma}{2} - i\Delta_{\rm ac} \right)t} \cosh(r)\sinh(r),            
\end{eqnarray}
where the corresponding variance of the acoustic displacement quadrature can be calculated as 
$\Delta^2 X_{\rm b}(t)=\bar{\mathcal{V}}_{{\rm b},11}(t)$. Thus the fidelity between state
$\mathcal{V}_{{\rm sg}}(0)$ and $\bar{\mathcal{V}}_{{\rm b}}(t)$ can be calculated as follows
\begin{eqnarray}
	\bar{\mathcal{F}} \left[ \mathcal{V}_{{\rm sg}}(0), \bar{\mathcal{V}}_{{\rm b}}(t)  \right]
	= \frac{1}{ { \sqrt{\Delta+\Lambda} - \sqrt{\Lambda} }},
\end{eqnarray}
with
\begin{eqnarray}
	\Delta &=& \det \left[ \mathcal{V}_{{\rm sg}}(0) + \bar{\mathcal{V}}_{{\rm b}}(t)  \right], \nonumber\\
	\Lambda &=& 4 \left[ \det \left( \mathcal{V}_{{\rm sg}}(0) - \frac{1}{4} \right) \right] 
	\left[ \det \left( \bar{\mathcal{V}}_{{\rm b}}(t) \right) - \frac{1}{4} \right].
\end{eqnarray}

After a storage period $\tau_{\rm s}$, we apply a second pump to the waveguide and transfer the state
from the acoustic phonons to retrieval photons. The state of the retrieval photons during the readou process 
can be described by the corresponding covariance matrix $\mathcal{V}_{\rm re}(t)$ with three independent elements as follows
\begin{eqnarray}
	\mathcal{V}_{{\rm re},11} &=& \frac{1}{2} - \frac{1}{2} e^{-2i\beta_{\rm re}}
	\tilde{\mu}_1^2 \left( e^{\tilde{\omega}_{+}t} - e^{\tilde{\omega}_{-}t} \right)^2
	\mu_1^2 \left( e^{\omega_{+}\tau_1} - e^{\omega_{-}\tau_1} \right)^2
	(2\bar{n}_{\rm th}+1) e^{2(-\frac{\Gamma}{2}+i\Delta_{\rm ac})\tau_{\rm s}} \cosh(r)\sinh(r) \nonumber\\
	&& - \frac{1}{2} e^{2i\beta_{\rm re}} \left( \tilde{\mu}_1^{*} \right)^2 
	\left( e^{\tilde{\omega}_{+}^{*}t} - e^{\tilde{\omega}_{-}^{*}t} \right)^2
	(\mu_1^{*})^2 \left( e^{\omega_{+}^{*}\tau_1} - e^{\omega_{-}^{*}\tau_1} \right)^2
	(2\bar{n}_{\rm th}+1) e^{2(-\frac{\Gamma}{2}-i\Delta_{\rm ac})\tau_{\rm s}} \cosh(r)\sinh(r) \nonumber\\
	&& + |\tilde{\mu}_1|^2 |e^{\tilde{\omega}_{+}t} - e^{\tilde{\omega}_{-}t}|^2
	\langle \tilde{b}^{\dagger}(0) \tilde{b}(0) \rangle \nonumber\\
	&& + \Gamma n_{\rm th} |\tilde{\mu}_1|^2 \left[ 
	\frac{1}{\tilde{\alpha}_1} \left( e^{\tilde{\alpha}_1t} - 1 \right)
	- \frac{1}{\tilde{\alpha}_2} \left( e^{\tilde{\alpha}_2t} - 1 \right)
	- \frac{1}{\tilde{\alpha}_3} \left( e^{\tilde{\alpha}_3t} - 1 \right) 
	+ \frac{1}{\tilde{\alpha}_4} \left( e^{\tilde{\alpha}_4t} - 1 \right) \right], \nonumber\\
	\mathcal{V}_{{\rm re},22} &=& \frac{1}{2} - \frac{1}{2} e^{-i(2\beta_{\rm re}+\pi)}	
	\tilde{\mu}_1^2 \left( e^{\tilde{\omega}_{+}t} - e^{\tilde{\omega}_{-}t} \right)^2
	\mu_1^2 \left( e^{\omega_{+}\tau_1} - e^{\omega_{-}\tau_1} \right)^2 
	(2\bar{n}_{\rm th}+1) e^{2(-\frac{\Gamma}{2}+i\Delta_{\rm ac})\tau_{\rm s}} \cosh(r)\sinh(r) \nonumber\\  
	&& - \frac{1}{2} e^{i(2\beta_{\rm re}+\pi)} \left( \tilde{\mu}_1^{*} \right)^2
	\left( e^{\tilde{\omega}_{+}^{*}t} - e^{\tilde{\omega}_{-}^{*}t} \right)^2
	(\mu_1^{*})^2 \left( e^{\omega_{+}^{*}\tau_1} - e^{\omega_{-}^{*}\tau_1} \right)^2
	(2\bar{n}_{\rm th}+1) e^{2(-\frac{\Gamma}{2}-i\Delta_{\rm ac})\tau_{\rm s}} \cosh(r)\sinh(r) \nonumber\\ 
	&& + |\tilde{\mu}_1|^2 |e^{\tilde{\omega}_{+}t} - e^{\tilde{\omega}_{-}t}|^2
	\langle \tilde{b}^{\dagger}(0) \tilde{b}(0) \rangle \nonumber\\  
	&& + \Gamma n_{\rm th} |\tilde{\mu}_1|^2 \left[ 
	\frac{1}{\tilde{\alpha}_1} \left( e^{\tilde{\alpha}_1t} - 1 \right)
	- \frac{1}{\tilde{\alpha}_2} \left( e^{\tilde{\alpha}_2t} - 1 \right)
	- \frac{1}{\tilde{\alpha}_3} \left( e^{\tilde{\alpha}_3t} - 1 \right)
	+ \frac{1}{\tilde{\alpha}_4} \left( e^{\tilde{\alpha}_4t} - 1 \right) \right], \nonumber\\
	\mathcal{V}_{{\rm re},12} &=& - \frac{1}{2} e^{-i(2\beta_{\rm re}+\pi/2)} \tilde{\mu}_1^2
	\left( e^{\tilde{\omega}_{+}t} - e^{\tilde{\omega}_{-}t} \right)^2	  
	\mu_1^2 \left( e^{\omega_{+}\tau_1} - e^{\omega_{-}\tau_1} \right)^2 
	(2\bar{n}_{\rm th}+1) e^{2(-\frac{\Gamma}{2}+i\Delta_{\rm ac})\tau_{\rm s}} \cosh(r)\sinh(r) \nonumber\\  
	&& - \frac{1}{2} e^{i(2\beta_{\rm re}+\pi/2)}  \left( \tilde{\mu}_1^{*} \right)^2 
	\left( e^{\tilde{\omega}_{+}^{*}t} - e^{\tilde{\omega}_{-}^{*}t} \right)^2  
	(\mu_1^{*})^2 \left( e^{\omega_{+}^{*}\tau_1} - e^{\omega_{-}^{*}\tau_1} \right)^2
	(2\bar{n}_{\rm th}+1) e^{2(-\frac{\Gamma}{2}-i\Delta_{\rm ac})\tau_{\rm s}} \cosh(r)\sinh(r), 
\end{eqnarray}
where coefficients $\tilde{\omega}_{\pm}$, $\tilde{\tau}_{\pm}$, $\tilde{\mu}_i$, and $\tilde{\alpha}_j$ are
listed in Eqs.~(\ref{Coefficents21}) and (\ref{Coefficents22}). The corresponding variance of the displacement
quadrature can be calculated as $\Delta^2 X_{\rm re}(t)=\mathcal{V}_{{\rm re},11}(t)$. 
The fidelity between the initial squeezed thermal state $\mathcal{V}_{\rm sg}(0)$ and the state 
$\mathcal{V}_{\rm re}(t)$ of retrieval photons can be written as follows
\begin{eqnarray}
	\tilde{\mathcal{F}}\left[ \mathcal{V}_{\rm sg}(0), \mathcal{V}_{\rm re}(t) \right]
	= \frac{1}{ { \sqrt{\Delta + \Lambda} - \sqrt{\Lambda} }},
\end{eqnarray}
with
\begin{eqnarray}
	\Delta &=& \det \left[  \mathcal{V}_{\rm sg}(0) + \mathcal{V}_{\rm re}(t) \right], \nonumber\\
	\Lambda &=& 4 \left[ \det\left[ \mathcal{V}_{\rm sg}(0) \right] - \frac{1}{4} \right]
	\left[ \det\left[ \mathcal{V}_{\rm re}(t) \right] - \frac{1}{4} \right].
\end{eqnarray}

%%%%%%%%%%%%%%%%%%%%%%%%%%%%%%%
%%%%%%%%%%%%%%%%%%%%%%%%%%%%%%%
\begin{figure}[t]
	\centerline{
		\includegraphics[clip, width=\linewidth]{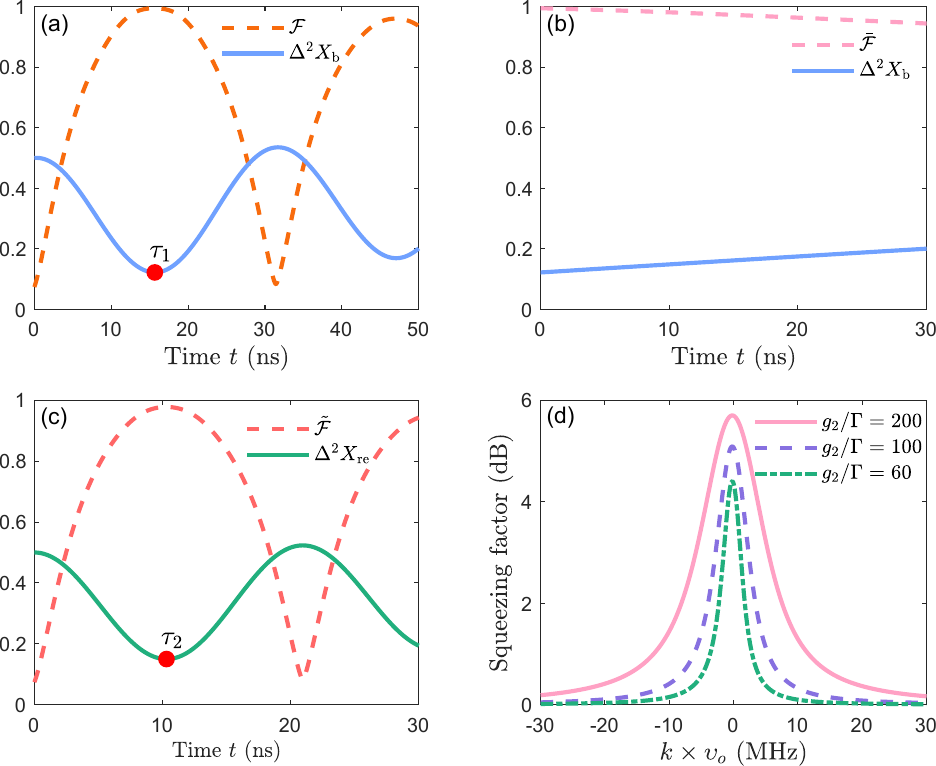}}
	\caption{ Simulation results for quantum memory of squeezed thermal states in Brillouin-active 
		waveguides environmental temperature $T_{\rm en}=1~{\rm K}$. 
		(a) Time evolution of quadrature variance $\Delta^2 X_{\rm b}$ and fidelity 
		$\mathcal{F}[\mathcal{V}_{sg}(0),\mathcal{V}_{\rm b}(t)]$ in the 
		writing process with coupling ratio $g_1/\Gamma=100$, where red point denotes the optimal
		time for maximum quantum transduction. (b) Time evolution of variance $\Delta^2 X_{\rm b}$
		and corresponding fidelity $\mathcal{F}[\mathcal{V}_{sg}(0),\bar{\mathcal{V}}_{\rm b}(t)]$
		during the storage process. Here, we switch off the first pump at the optimal time $t=\tau_1$
		in the writing process. (c) Time evolution of $\Delta^2 X_{\rm re}$
		and $\tilde{\mathcal{F}}[\mathcal{V}_{sg}(0),\mathcal{V}_{\rm re}(t)]$ in the readout process
		with a coupling ratio $g_2/\Gamma=150$ and a storage time of $\tau_{\rm s}=5~{\rm ns}$.
		(d) Continuum memory versus the wave number $k$ in the strong coupling regime ($g_1=g_2$).  
	}\label{Sup_fig1}
\end{figure}
%%%%%%%%%%%%%%%%%%%%%%%%%%%%%%%
%%%%%%%%%%%%%%%%%%%%%%%%%%%%%%%

\subsection{Memory of squeezed coherent states}
In this subsection, we investigate the quantum memory for the squeezed coherent state in 
Brillouin-active waveguides. Here, we assume that the signal photons are prepared to
a squeezed coherent state as $|r,\alpha\rangle=S(r)|\alpha\rangle$, where $S(r)$ represents the
unitary phase-free squeezing operator and $|\alpha\rangle$ denotes a coherent state. 
The corresponding covariance matrix of the initial signal photons is
\begin{eqnarray}
	\mathcal{V}_{\rm sg}(0) = \left[
	\begin{array}{cc}
		\frac{1}{2} e^{-2r} & 0 \\
		0 & \frac{1}{2} e^{2r}
	\end{array}
	\right],
\end{eqnarray}
which is similar to the covariance matrix of a squeezed vacuum state.
During the writing process, the symmetric covariance matrix $\mathcal{V}_{\rm b}(t)$ of the acoustic phonons
can be characterized as follows
\begin{eqnarray}
	\mathcal{V}_{{\rm b},11}(t) &=& \frac{1}{2} - \frac{1}{2} e^{-2i\beta_b} \mu_1^2 
	\left( e^{\omega_{+}t} - e^{\omega_{-}t} \right)^2 \cosh(r)\sinh(r) \nonumber\\
	&& - \frac{1}{2} e^{2i\beta_b} (\mu_1^{*})^2
	\left( e^{\omega_{+}^{*}t} - e^{\omega_{-}^{*}t} \right)^2 \cosh(r)\sinh(r) \nonumber\\
	&& + |\mu_1|^2 \left| e^{\omega_{+}t} - e^{\omega_{-}t} \right|^2 \sinh^2(r) \nonumber\\
	&& + \Gamma n_{\rm th} \left[ 
	\frac{|\mu_3|^2}{\alpha_1} \left( e^{\alpha_1t} - 1 \right)
	- \frac{\mu_3\mu_2^{*}}{\alpha_2} \left( e^{\alpha_2t} - 1 \right)
	- \frac{\mu_2\mu_3^{*}}{\alpha_3} \left( e^{\alpha_3t} - 1 \right) 
	+ \frac{|\mu_2|^2}{\alpha_1} \left( e^{\alpha_4t} - 1 \right) \right], \nonumber\\
	\mathcal{V}_{{\rm b},22}(t) &=&	 \frac{1}{2} - \frac{1}{2} e^{-i(2\beta_b+\pi)} \mu_1^2 
	\left( e^{\omega_{+}t} - e^{\omega_{-}t} \right)^2 \cosh(r)\sinh(r) \nonumber\\   
	&& - \frac{1}{2} e^{i(2\beta_b+\pi)} (\mu_1^{*})^2  
	\left( e^{\omega_{+}^{*}t} - e^{\omega_{-}^{*}t} \right)^2 \cosh(r)\sinh(r) \nonumber\\ 
	&& + |\mu_1|^2 \left| e^{\omega_{+}t} - e^{\omega_{-}t} \right|^2 \sinh^2(r) \nonumber\\   
	&& + \Gamma n_{\rm th} \left[ 
	\frac{|\mu_3|^2}{\alpha_1} \left( e^{\alpha_1t} - 1 \right)
	- \frac{\mu_3\mu_2^{*}}{\alpha_2} \left( e^{\alpha_2t} - 1 \right)
	- \frac{\mu_2\mu_3^{*}}{\alpha_3} \left( e^{\alpha_3t} - 1 \right)
	+ \frac{|\mu_2|^2}{\alpha_1} \left( e^{\alpha_4t} - 1 \right) \right], \nonumber\\
	\mathcal{V}_{{\rm b},22}(t) &=&	- \frac{1}{2} e^{-i(2\beta_b+\pi/2)} \mu_1^2
	\left( e^{\omega_{+}t} - e^{\omega_{-}t} \right)^2 \cosh(r)\sinh(r) \nonumber\\  
	&&  - \frac{1}{2} e^{i(2\beta_b+\pi/2)} (\mu_1^{*})^2
	\left( e^{\omega_{+}^{*}t} - e^{\omega_{-}^{*}t} \right)^2 \cosh(r)\sinh(r)  ,        
\end{eqnarray}
where coefficients $\omega_{\pm}$, $\tau_{\pm}$, $\mu_{i}$, and $\alpha_{j}$ are described by 
Eqs.~(\ref{Coefficients11}) and (\ref{Coefficients12}). The variance of the displacement quadrature
of acoustic phonons in the writing process can be evaluated by $\Delta^2 X_{\rm b}(t)=\mathcal{V}_{{\rm b},11}(t)$.
The fidelity between the initial state of signal photons and the state of acoustic phonons at time $t$
can be calculated as follows
\begin{eqnarray}
	\mathcal{F} \left[ \mathcal{V}_{\rm sg}(0), \mathcal{V}_{\rm b}(t) \right]
	&=& \frac{1}{ { \sqrt{\Delta+\Lambda} - \sqrt{\Lambda} }}
	\exp \left[ -\frac{1}{4} \delta_u^T 
	\left(  \mathcal{V}_{\rm sg}(0) + \mathcal{V}_{\rm b}(t) \right)^{-1} \delta_u \right],
\end{eqnarray}
where
\begin{eqnarray}
	\Delta &=& \det \left[ \mathcal{V}_{\rm sg}(0) + \mathcal{V}_{\rm b}(t) \right], \nonumber\\
	\Lambda &=& 4\left[ \det\left[ \mathcal{V}_{\rm sg}(0) \right] - \frac{1}{4} \right]
	\left[ \det\left[ \mathcal{V}_{\rm b}(t) \right] - \frac{1}{4} \right], \nonumber\\
	\delta_u &=& \left[ 
	\begin{array}{cc}
		u_1 \\
		u_2
	\end{array} \right],
\end{eqnarray}
and
\begin{eqnarray}
	u_1(t) &=& \frac{1}{\sqrt{2}} \left[ 1 - e^{-i\beta_b} \mu_1 \left(  e^{\omega_{+}t} - e^{\omega_{-}t} \right) \right]
	\left( \alpha\cosh(r) - \alpha^{*}\sinh(r) \right) \nonumber\\
	&& + \frac{1}{\sqrt{2}} \left[ 1 - e^{i\beta_b} \mu_1^{*} \left(  e^{\omega_{+}^{*}t} - e^{\omega_{-}^{*}t} \right) \right] 
	\left( \alpha^{*}\cosh(r) - \alpha\sinh(r) \right), \nonumber\\
	u_2(t) &=& -\frac{1}{\sqrt{2}} \left[ i + e^{-i(\beta_b+\pi/2)} \mu_1 \left(  e^{\omega_{+}t} - e^{\omega_{-}t} \right) \right]
	\left( \alpha\cosh(r) - \alpha^{*}\sinh(r) \right) \nonumber\\
	&& + \frac{1}{\sqrt{2}} \left[ i - e^{i(\beta_b+\pi/2)} \mu_1^{*} \left(  e^{\omega_{+}^{*}t} - e^{\omega_{-}^{*}t} \right) \right]
	\left( \alpha\cosh(r) - \alpha^{*}\sinh(r) \right).
\end{eqnarray}
As the variance $\Delta^2 X_{\rm b}(t)$ during the writing process experiences a Rabi oscillation, 
its minimum value can be achieved at the optical time $\tau_1\approx\pi/(2g_1)$, where we choose phase $\beta_b=3\pi/2$.

In the storage process, the acoustic phonons are driven by the thermal noise and thereby the corresponding covariance matrix
$\bar{\mathcal{V}}_{\rm b}(t)$ can be characterized as
\begin{eqnarray}
	\bar{\mathcal{V}}_{{\rm b},11}(t) &=& \frac{1}{2} - \frac{1}{2} e^{-2i\beta_b} \mu_1^2
	\left( e^{\omega_{+}\tau_1} - e^{\omega_{-}\tau_1} \right)^2
	e^{2\left( -\frac{\Gamma}{2}+i\Delta_{\rm ac} \right)t} \cosh(r)\sinh(r) \nonumber\\
	&& - \frac{1}{2} e^{2i\beta_b} (\mu_1^{*})^2
	\left( e^{\omega_{+}^{*}\tau_1} - e^{\omega_{-}^{*}\tau_1} \right)^2
	e^{2\left( -\frac{\Gamma}{2}-i\Delta_{\rm ac} \right)t} \cosh(r)\sinh(r) \nonumber\\
	&& + n_{\rm th} \left( 1 - e^{-\Gamma t} \right)
	+ e^{-\Gamma t}|\mu_1|^2 \left| e^{\omega_{+}\tau_1} - e^{\omega_{-}\tau_1} \right|^2 \sinh^2(r) \nonumber\\
	&& + e^{-\Gamma t} \Gamma n_{\rm th} \left[ 
	\frac{|\mu_3|^2}{\alpha_1} \left( e^{\alpha_1\tau_1} - 1 \right)
	- \frac{\mu_3\mu_2^{*}}{\alpha_2} \left( e^{\alpha_2\tau_1} - 1 \right)
	- \frac{\mu_2\mu_3^{*}}{\alpha_3} \left( e^{\alpha_3\tau_1} - 1 \right)
	+ \frac{|\mu_2|^2}{\alpha_4} \left( e^{\alpha_4\tau_1} - 1 \right) \right], \nonumber\\
	\bar{\mathcal{V}}_{{\rm b},22}(t) &=& \frac{1}{2} - \frac{1}{2} e^{-i(2\beta_b+\pi)} \mu_1^2  
	\left( e^{\omega_{+}\tau_1} - e^{\omega_{-}\tau_1} \right)^2 
	e^{2\left( -\frac{\Gamma}{2}+i\Delta_{\rm ac} \right)t} \cosh(r)\sinh(r) \nonumber\\ 
	&& - \frac{1}{2} e^{i(2\beta_b+\pi)} (\mu_1^{*})^2 
	\left( e^{\omega_{+}^{*}\tau_1} - e^{\omega_{-}^{*}\tau_1} \right)^2 
	e^{2\left( -\frac{\Gamma}{2}-i\Delta_{\rm ac} \right)t} \cosh(r)\sinh(r) \nonumber\\  
	&& + n_{\rm th} \left( 1 - e^{-\Gamma t} \right)
	+ e^{-\Gamma t}|\mu_1|^2 \left| e^{\omega_{+}\tau_1} - e^{\omega_{-}\tau_1} \right|^2 \sinh^2(r) \nonumber\\  
	&& + e^{-\Gamma t} \Gamma n_{\rm th} \left[
	\frac{|\mu_3|^2}{\alpha_1} \left( e^{\alpha_1\tau_1} - 1 \right)
	- \frac{\mu_3\mu_2^{*}}{\alpha_2} \left( e^{\alpha_2\tau_1} - 1 \right)
	- \frac{\mu_2\mu_3^{*}}{\alpha_3} \left( e^{\alpha_3\tau_1} - 1 \right)
	+ \frac{|\mu_2|^2}{\alpha_4} \left( e^{\alpha_4\tau_1} - 1 \right) \right], \nonumber\\ 
	\bar{\mathcal{V}}_{{\rm b},12}(t) &=& - \frac{1}{2} e^{-i(2\beta_b+\pi/2)}   
	e^{2\left( -\frac{\Gamma}{2}+i\Delta_{\rm ac} \right)t} \mu_1^2  
	\left( e^{\omega_{+}\tau_1} - e^{\omega_{-}\tau_1} \right)^2  
	\cosh(r)\sinh(r) \nonumber\\ 
	&& - \frac{1}{2} e^{i(2\beta_b+\pi/2)} 
	e^{2\left( -\frac{\Gamma}{2}-i\Delta_{\rm ac} \right)t} (\mu_1^{*})^2  
	\left( e^{\omega_{+}^{*}\tau_1} - e^{\omega_{-}^{*}\tau_1} \right)^2 
	\cosh(r)\sinh(r).                    
\end{eqnarray}
The variance of the acoustic displacement quadrature can be calculated as 
$\Delta^2 X_{\rm b}(t)=\bar{\mathcal{V}}_{{\rm b},11}(t)$. The corresponding fidelity
between the initial state of signal photons and the state of acoustic phonons during the 
storage process can be given by
\begin{eqnarray}
	\bar{\mathcal{F}} \left[ \mathcal{V}_{\rm sg}(0), \bar{\mathcal{V}}_{\rm b}(t) \right]
	&=& \frac{1}{ { \sqrt{\Delta+\Lambda} - \sqrt{\Lambda} }}
	\exp \left[ -\frac{1}{2}\delta_u^T 
	\left( \mathcal{V}_{\rm sg}(0) + \bar{\mathcal{V}}_{\rm b}(t) \right)^{-1} \delta_u \right],
\end{eqnarray}
where
\begin{eqnarray}
	\Delta &=& \det \left[ \mathcal{V}_{\rm sg}(0) + \bar{\mathcal{V}}_{\rm b}(t) \right], \nonumber\\
	\Lambda &=& 4 \left[ \det\left[ \mathcal{V}_{\rm sg}(0) \right] - \frac{1}{4} \right]
	\left[ \det\left[ \bar{\mathcal{V}}_{\rm b}(t) \right] - \frac{1}{4} \right], \nonumber\\
	\delta_u &=& \left[ 
	\begin{array}{cc}
		\bar{u}_1 \\
		\bar{u}_2
	\end{array} \right],
\end{eqnarray}
and
\begin{eqnarray}
	\bar{u}_1(t) &=& \frac{1}{\sqrt{2}} \left[ \alpha\cosh(r) - \alpha^{*}\sinh(r) \right] 
	+ \frac{1}{\sqrt{2}} \left[ \alpha^{*}\cosh(r) - \alpha\sinh(r) \right] \nonumber\\
	&&  - \frac{1}{\sqrt{2}} e^{-i\beta_b} e^{(-\frac{\Gamma}{2}+i\Delta_{\rm ac})t}
	\mu_1 \left( e^{\omega_{+}\tau_1} - e^{\omega_{-}\tau_1} \right)
	\left[ \alpha\cosh(r) - \alpha^{*}\sinh(r) \right] \nonumber\\
	&&  - \frac{1}{\sqrt{2}} e^{i\beta_b} e^{(-\frac{\Gamma}{2}-i\Delta_{\rm ac})t}
	\mu_1^{*} \left( e^{\omega_{+}^{*}\tau_1} - e^{\omega_{-}^{*}\tau_1} \right)
	\left[ \alpha^{*}\cosh(r) - \alpha\sinh(r) \right], \nonumber \\
	\bar{u}_2(t) &=& -\frac{i}{\sqrt{2}} \left[ \alpha\cosh(r) - \alpha^{*}\sinh(r) \right]	
	+ \frac{i}{\sqrt{2}} \left[ \alpha^{*}\cosh(r) - \alpha\sinh(r) \right] \nonumber\\  
	&&  - \frac{1}{\sqrt{2}} e^{-i(\beta_b+\pi/2)} e^{(-\frac{\Gamma}{2}+i\Delta_{\rm ac})t}
	\mu_1 \left( e^{\omega_{+}\tau_1} - e^{\omega_{-}\tau_1} \right)
	\left[ \alpha\cosh(r) - \alpha^{*}\sinh(r) \right] \nonumber\\  
	&&  - \frac{1}{\sqrt{2}} e^{i(\beta_b+\pi/2)} e^{(-\frac{\Gamma}{2}-i\Delta_{\rm ac})t} 
	\mu_1^{*} \left( e^{\omega_{+}^{*}\tau_1} - e^{\omega_{-}^{*}\tau_1} \right)  
	\left[ \alpha^{*}\cosh(r) - \alpha\sinh(r) \right].         
\end{eqnarray}

After a storage period $\tau_{\rm s}$, we apply a second pump to the waveguide and transfer the
squeezed state from acoustic phonons to retrieval photons. In this readout process, the symmetric
covariance matrix $\mathcal{V}_{\rm re}$ of the retrieval photons can be expressed as
\begin{eqnarray}
	\mathcal{V}_{{\rm re},11}(t) &=& \frac{1}{2} - \frac{1}{2} e^{-2i\beta_{\rm re}}
	\tilde{\mu}_1^2 \left( e^{\tilde{\omega}_{+}t} - e^{\tilde{\omega}_{-}t} \right)^2
	e^{2(-\frac{\Gamma}{2}+i\Delta_{\rm ac})\tau_{\rm s}} \mu_1^2
	\left( e^{\omega_{+}\tau_1} - e^{\omega_{-}\tau_1} \right)^2
	\cosh(r)\sinh(r) \nonumber\\
	&& - \frac{1}{2} e^{2i\beta_{\rm re}} (\tilde{\mu}_1^{*})^2
	\left( e^{\tilde{\omega}_{+}^{*}t} - e^{\tilde{\omega}_{-}^{*}t} \right)^2 
	e^{2(-\frac{\Gamma}{2}-i\Delta_{\rm ac})\tau_{\rm s}} (\mu_1^{*})^2
	\left( e^{\omega_{+}^{*}\tau_1} - e^{\omega_{-}^{*}\tau_1} \right)^2 
	\cosh(r)\sinh(r) \nonumber\\
	&& + |\tilde{\mu}_1|^2 \left| e^{\tilde{\omega}_{+}t} - e^{\tilde{\omega}_{-}t} \right|^2
	\langle \tilde{b}^{\dagger}(0) \tilde{b}(0) \rangle \nonumber\\
	&& + \Gamma n_{\rm th} |\tilde{\mu}_1|^2 \left[ 
	\frac{1}{\tilde{\alpha}_1} \left( e^{\tilde{\alpha}_1t} - 1 \right)
	- \frac{1}{\tilde{\alpha}_2} \left( e^{\tilde{\alpha}_2t} - 1 \right)
	- \frac{1}{\tilde{\alpha}_3} \left( e^{\tilde{\alpha}_3t} - 1 \right)
	+ \frac{1}{\tilde{\alpha}_4} \left( e^{\tilde{\alpha}_4t} - 1 \right)  \right], \nonumber\\
	\mathcal{V}_{{\rm re},22}(t) &=& \frac{1}{2} - \frac{1}{2} e^{-2i(\beta_{\rm re}+\pi/2)}
	\tilde{\mu}_1^2 \left( e^{\tilde{\omega}_{+}t} - e^{\tilde{\omega}_{-}t} \right)^2
	e^{2(-\frac{\Gamma}{2}+i\Delta_{\rm ac})\tau_{\rm s}} \mu_1^2	
	\left( e^{\omega_{+}\tau_1} - e^{\omega_{-}\tau_1} \right)^2  
	\cosh(r)\sinh(r) \nonumber\\  
	&& - \frac{1}{2} e^{2i(\beta_{\rm re}+\pi/2)} (\tilde{\mu}_1^{*})^2
	\left( e^{\tilde{\omega}_{+}^{*}t} - e^{\tilde{\omega}_{-}^{*}t} \right)^2 
	e^{2(-\frac{\Gamma}{2}-i\Delta_{\rm ac})\tau_{\rm s}} (\mu_1^{*})^2 
	\left( e^{\omega_{+}^{*}\tau_1} - e^{\omega_{-}^{*}\tau_1} \right)^2 
	\cosh(r)\sinh(r) \nonumber\\  
	&& + |\tilde{\mu}_1|^2 \left| e^{\tilde{\omega}_{+}t} - e^{\tilde{\omega}_{-}t} \right|^2
	\langle \tilde{b}^{\dagger}(0) \tilde{b}(0) \rangle \nonumber\\ 
	&& + \Gamma n_{\rm th} |\tilde{\mu}_1|^2 \left[ 
	\frac{1}{\tilde{\alpha}_1} \left( e^{\tilde{\alpha}_1t} - 1 \right)
	- \frac{1}{\tilde{\alpha}_2} \left( e^{\tilde{\alpha}_2t} - 1 \right)
	- \frac{1}{\tilde{\alpha}_3} \left( e^{\tilde{\alpha}_3t} - 1 \right)
	+ \frac{1}{\tilde{\alpha}_4} \left( e^{\tilde{\alpha}_4t} - 1 \right)  \right], \nonumber\\   
	\mathcal{V}_{{\rm re},12}(t) &=& - \frac{1}{2} e^{-i(2\beta_{\rm re}+\pi/2)}
	\tilde{\mu}_1^2 \left( e^{\tilde{\omega}_{+}t} - e^{\tilde{\omega}_{-}t} \right)^2	 
	e^{2(-\frac{\Gamma}{2}+i\Delta_{\rm ac})\tau_{\rm s}} \mu_1^2	
	\left( e^{\omega_{+}\tau_1} - e^{\omega_{-}\tau_1} \right)^2 
	\cosh(r)\sinh(r) \nonumber\\     
	&& - \frac{1}{2} e^{2i(\beta_{\rm re}+\pi/2)} (\tilde{\mu}_1^{*})^2 
	\left( e^{\tilde{\omega}_{+}^{*}t} - e^{\tilde{\omega}_{-}^{*}t} \right)^2
	e^{2(-\frac{\Gamma}{2}-i\Delta_{\rm ac})\tau_{\rm s}} (\mu_1^{*})^2
	\left( e^{\omega_{+}^{*}\tau_1} - e^{\omega_{-}^{*}\tau_1} \right)^2
	\cosh(r)\sinh(r),  
\end{eqnarray}
with
\begin{eqnarray}
	\langle \tilde{b}^{\dagger}(0) \tilde{b}(0) \rangle &=& n_{\rm th} \left( 1 - e^{-\Gamma\tau_{\rm s}} \right)
	+ e^{-\Gamma\tau_{\rm s}} |\mu_1|^2 \left| e^{\omega_{+}\tau_1} - e^{\omega_{-}\tau_1} \right|^2 \sinh^2(r) \nonumber\\
	&& + e^{-\Gamma\tau_{\rm s}} \Gamma n_{\rm th} \left[ 
	\frac{|\mu_3|^2}{\alpha_1} \left( e^{\alpha_1\tau_1} - 1 \right)
	- \frac{\mu_3\mu_2^{*}}{\alpha_2} \left( e^{\alpha_2\tau_1} - 1 \right)
	- \frac{\mu_2\mu_3^{*}}{\alpha_3} \left( e^{\alpha_3\tau_1} - 1 \right)
	+ \frac{|\mu_2|^2}{\alpha_4} \left( e^{\alpha_4\tau_1} - 1 \right) \right]
\end{eqnarray}
The variance of the acoustic displacement quadrature can be calculated as 
$\Delta^2 X_{\rm re}(t)=\mathcal{V}_{{\rm re},11}(t)$. The fidelity between the initial state of signal photons
and the state of the retrieval photons can be given by
\begin{eqnarray}
	\tilde{\mathcal{F}}\left[ \mathcal{V}_{\rm sg}(0), \mathcal{V}_{\rm re}(t) \right] &=&
	\frac{1}{ { \sqrt{\Delta+\Lambda} - \sqrt{\Lambda} }}
	\exp \left[ -\frac{1}{2}\delta_u^T \left( \mathcal{V}_{\rm sg}(0) + \mathcal{V}_{\rm re}(t) \right)^{-1} \delta_u \right],
\end{eqnarray}
with
\begin{eqnarray}
	\Delta &=& \det\left[ \mathcal{V}_{\rm sg}(0) + \mathcal{V}_{\rm re}(t) \right], \nonumber\\
	\Lambda &=& 4 \left[ \det\left[ \mathcal{V}_{\rm sg}(0) \right] - \frac{1}{4} \right]
	\left[ \det\left[ \mathcal{V}_{\rm re}(t) \right] - \frac{1}{4} \right], \nonumber\\
	\delta_u &=& \left[ 
	\begin{array}{cc}
		\tilde{u}_1 \\
		\tilde{u}_2
	\end{array} \right],
\end{eqnarray}
and
\begin{eqnarray}
	\tilde{u}_1 &=& \frac{1}{\sqrt{2}} \left[ \alpha\cosh(r) -\alpha^{*}\sinh(r) \right]
	+ \frac{1}{\sqrt{2}} \left[ \alpha^{*}\cosh(r) -\alpha\sinh(r) \right] \nonumber\\
	&& - \frac{1}{\sqrt{2}} e^{-i\beta_{\rm re}} \tilde{\mu}_1 
	\left( e^{\tilde{\omega}_{+}t} - e^{\tilde{\omega}_{-}t} \right)
	e^{(-\frac{\Gamma}{2}+i\Delta_{\rm ac})\tau_{\rm s}} \mu_1
	\left( e^{\omega_{+}\tau_1} - e^{\omega_{-}\tau_1} \right) 
	\left[ \alpha\cosh(r) -\alpha^{*}\sinh(r) \right] \nonumber\\
	&& - \frac{1}{\sqrt{2}} e^{i\beta_{\rm re}} \tilde{\mu}_1^{*}
	\left( e^{\tilde{\omega}_{+}^{*}t} - e^{\tilde{\omega}_{-}^{*}t} \right)
	e^{(-\frac{\Gamma}{2}-i\Delta_{\rm ac})\tau_{\rm s}} \mu_1^{*}
	\left( e^{\omega_{+}^{*}\tau_1} - e^{\omega_{-}^{*}\tau_1} \right)
	\left[ \alpha^{*}\cosh(r) -\alpha\sinh(r) \right], \nonumber\\
	\tilde{u}_2 &=&	- \frac{i}{\sqrt{2}} \left[ \alpha\cosh(r) -\alpha^{*}\sinh(r) \right]  
	+ \frac{i}{\sqrt{2}} \left[ \alpha^{*}\cosh(r) -\alpha\sinh(r) \right] \nonumber\\  
	&& - \frac{1}{\sqrt{2}} e^{-i(\beta_{\rm re}+\pi/2)} \tilde{\mu}_1   
	\left( e^{\tilde{\omega}_{+}t} - e^{\tilde{\omega}_{-}t} \right)
	e^{(-\frac{\Gamma}{2}+i\Delta_{\rm ac})\tau_{\rm s}} \mu_1
	\left( e^{\omega_{+}\tau_1} - e^{\omega_{-}\tau_1} \right)  
	\left[ \alpha\cosh(r) -\alpha^{*}\sinh(r) \right] \nonumber\\ 
	&& - \frac{1}{\sqrt{2}} e^{i(\beta_{\rm re}+\pi/2)} \tilde{\mu}_1^{*} 
	\left( e^{\tilde{\omega}_{+}^{*}t} - e^{\tilde{\omega}_{-}^{*}t} \right) 
	e^{(-\frac{\Gamma}{2}-i\Delta_{\rm ac})\tau_{\rm s}} \mu_1^{*} 
	\left( e^{\omega_{+}^{*}\tau_1} - e^{\omega_{-}^{*}\tau_1} \right) 
	\left[ \alpha^{*}\cosh(r) -\alpha\sinh(r) \right].   
\end{eqnarray}

%%%%%%%%%%%%%%%%%%%%%%%%%%%%%%%
%%%%%%%%%%%%%%%%%%%%%%%%%%%%%%%
\begin{figure}[t]
	\centerline{
		\includegraphics[clip, width=\linewidth]{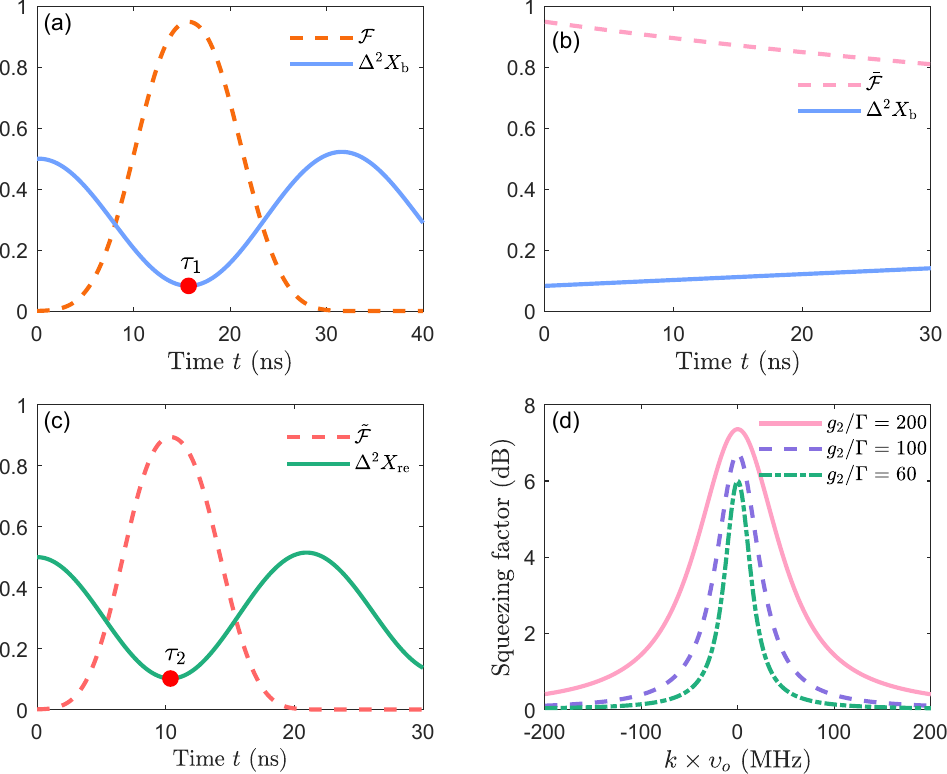}}
	\caption{ Simulation results for quantum memory of squeezed coherent states in Brillouin-active 
		waveguides environmental temperature $T_{\rm en}=0.7~{\rm K}$. 
		(a) Time evolution of quadrature variance $\Delta^2 X_{\rm b}$ and fidelity 
		$\mathcal{F}[\mathcal{V}_{sg}(0),\mathcal{V}_{\rm b}(t)]$ in the 
		writing process with coupling ratio $g_1/\Gamma=100$, where red point denotes the optimal
		time for maximum quantum transduction. (b) Time evolution of variance $\Delta^2 X_{\rm b}$
		and corresponding fidelity $\mathcal{F}[\mathcal{V}_{sg}(0),\bar{\mathcal{V}}_{\rm b}(t)]$
		during the storage process. Here, we switch off the first pump at the optimal time $t=\tau_1$
		in the writing process. (c) Time evolution of $\Delta^2 X_{\rm re}$
		and $\tilde{\mathcal{F}}[\mathcal{V}_{sg}(0),\mathcal{V}_{\rm re}(t)]$ in the readout process
		with a coupling ratio $g_2/\Gamma=150$ and a storage time of $\tau_{\rm s}=5~{\rm ns}$.
		(d) Continuum memory versus the wave number $k$ in the strong coupling regime ($g_1=g_2$).  
	}\label{Sup_fig2}
\end{figure}
%%%%%%%%%%%%%%%%%%%%%%%%%%%%%%%
%%%%%%%%%%%%%%%%%%%%%%%%%%%%%%%

\section{Quantum memory for entangled states}\label{S3}
In this section, we discuss the quantum memory of optical entangled state in Brillouin-active waveguides.
Here we consider the pair of entangled lights, including a signal light $a_{\rm re}(t,k)$ and an idler light
$a_{\rm id}(t,k)$, which can be realized through a parametric-down conversion in a high nonlinear material.
$a_{\rm re}(t,k)$ ($a_{\rm id}(t,k)$) denotes the annihilation operator of the $k$-th signal (idler) mode with
wave number $k$. The state of signal and idler photons is a two-mode Gaussian state, which can be characterized
by its covariance matrix. We define the quadrature operators of the signal and idler photons as follows
\begin{eqnarray}
	X_{\rm sg} &=& \frac{ a_{\rm sg} + a_{\rm sg}^{\dagger} }{\sqrt{2}}, \quad
	P_{\rm sg} = \frac{ a_{\rm sg} - a_{\rm sg}^{\dagger} }{i\sqrt{2}}, \nonumber\\
	X_{\rm re} &=& \frac{ a_{\rm re} + a_{\rm re}^{\dagger} }{\sqrt{2}}, \quad
	P_{\rm re} = \frac{ a_{\rm re} - a_{\rm re}^{\dagger} }{i\sqrt{2}}.
\end{eqnarray}
Considering a simple case, the symmetric covariance matrix at the initial time can be given by
\begin{eqnarray}
	\mathcal{V}_{\rm re, id}(t=0) = \left[
	\begin{array}{cccc}
		\frac{1}{2}+2\eta^2 & 0 & -2\eta^2 & -\eta \\
		0 & \frac{1}{2}+2\eta^2 & -\eta & 2\eta^2 \\
		-2\eta^2 & -\eta & \frac{1}{2}+2\eta^2 & 0 \\
		-\eta & 2\eta^2 & 0 & \frac{1}{2}+2\eta^2 \\
	\end{array} \right].
\end{eqnarray}
Here $\eta = G_0/\gamma_0$, where $G_0$ denotes the effective coupling strength related 
to the pump power and nonlinearity of the material during the parametric-down conversion 
and $\gamma_0$ represents the optical damping rate in the nonlinear material. 
We feed the idler light into a single mode fiber for reference and feed the signal light
to a Brillouin-active waveguide for memory.

During the writing process, the dynamics of our system can be given by
\begin{eqnarray}\label{Dynamics of entanglement memory in writing process}
	\frac{d a_{\rm id}}{dt} &=& \left( -\frac{\gamma_{\rm smf}}{2} + i\Delta_{\rm id} \right) a_{\rm id}
	+ \sqrt{\gamma_{\rm smf}} \xi_{\rm id}, \nonumber \\
	\frac{d a_{\rm sg}}{dt} &=& \left( -\frac{\gamma}{2} + i\Delta_{\rm sg} \right) a_{\rm sg}
	-ig_1 b + \sqrt{\gamma}\xi_{\rm sg}, \nonumber\\
	\frac{d b}{dt} &=& \left( -\frac{\Gamma}{2} + i\Delta_{\rm ac} \right)b - ig_1 a_{\rm sg} + \sqrt{\Gamma}\xi_{\rm ac},
\end{eqnarray}
where we assume that the idler light is driven by the Langevin noise $\xi_{\rm id}$ in a 
single mode fiber with optical damping rate $\gamma_{\rm smf}$. 
By solving Eqs.~(\ref{Dynamics of entanglement memory in writing process}), we can obtain the
analytical solution of the covariance matrix $\mathcal{V}_{\rm id,b}$ as follows
\begin{eqnarray}
	\mathcal{V}_{\rm id, b}(t) = \left[
	\begin{array}{cccc}
		\mathcal{V}_{{\rm id,b},11} &                           0 & \mathcal{V}_{{\rm id,b},13} &  \mathcal{V}_{{\rm id,b},14} \\
		0                           & \mathcal{V}_{{\rm id,b},11} & \mathcal{V}_{{\rm id,b},14} & -\mathcal{V}_{{\rm id,b},13} \\
		\mathcal{V}_{{\rm id,b},13} & \mathcal{V}_{{\rm id,b},14} & \mathcal{V}_{{\rm id,b},33} &  0 \\
		\mathcal{V}_{{\rm id,b},14} &-\mathcal{V}_{{\rm id,b},13} & 0                           & \mathcal{V}_{{\rm id,b},33} \\
	\end{array} \right],
\end{eqnarray}
where these four independent elements can be expressed as
\begin{eqnarray}
	\mathcal{V}_{{\rm id,b},11}(t) &=& \frac{1}{2} + 2\eta^2 e^{\left( \lambda+\lambda^{*} \right)t}, \nonumber\\
	\mathcal{V}_{{\rm id,b},33}(t) &=& \frac{1}{2} + 2\eta^2 \left| \mu_1\left( e^{\omega_{+}t} - e^{\omega_{-}t} \right) \right|^2 \nonumber\\
	&& + \Gamma n_{\rm th} \left[ 
	\frac{|\mu_3|^2}{\alpha_1} \left( e^{\alpha_1t} - 1 \right)
	- \frac{\mu_3\mu_2^{*}}{\alpha_2} \left( e^{\alpha_2t} - 1 \right)
	- \frac{\mu_2\mu_3^{*}}{\alpha_3} \left( e^{\alpha_3t} - 1 \right)
	+ \frac{|\mu_2|^2}{\alpha_4} \left( e^{\alpha_4t} - 1 \right) \right], \nonumber\\
	\mathcal{V}_{{\rm id,b},13}(t) &=& \frac{i}{2}\mu_1 e^{\lambda_1t}	
	\left( e^{\omega_{+}t} - e^{\omega_{-}t} \right) \left( -2\eta^2-i\eta \right)  
	-\frac{i}{2}\mu_1^{*} e^{\lambda_1^{*}t} 
	\left( e^{\omega_{+}^{*}t} - e^{\omega_{-}^{*}t} \right) \left( -2\eta^2+i\eta \right) \nonumber\\
	\mathcal{V}_{{\rm id,b},14}(t) &=&	\frac{1}{2}\mu_1 e^{\lambda_1t}  
	\left( e^{\omega_{+}t} - e^{\omega_{-}t} \right) \left( -2\eta^2-i\eta \right)   
	+ \frac{1}{2}\mu_1^{*} e^{\lambda_1^{*}t}   
	\left( e^{\omega_{+}^{*}t} - e^{\omega_{-}^{*}t} \right) \left( -2\eta^2+i\eta \right),          
\end{eqnarray}
with coefficients
\begin{eqnarray}
	\omega_{+} &=& - \frac{\Gamma+\gamma}{4} + i\frac{\Delta_{\rm sg}+\Delta_{\rm ac}}{2} 
	- i\frac{ \sqrt{ 16g_1^2 - \left[ (\Gamma-\gamma) + 2i(\Delta_{\rm sg}-\Delta_{\rm ac}) \right]^2 } }{4}, \nonumber\\
	\omega_{-} &=& - \frac{\Gamma+\gamma}{4} + i\frac{\Delta_{\rm sg}+\Delta_{\rm ac}}{2}
	+ i\frac{ \sqrt{ 16g_1^2 - \left[ (\Gamma-\gamma) + 2i(\Delta_{\rm sg}-\Delta_{\rm ac}) \right]^2 } }{4}, \nonumber\\	
	\tau_{+}   &=& \frac{ -2(\Delta_{\rm sg}-\Delta_{\rm ac}) + i(\Gamma-\gamma) }{4g_1}
	+ \frac{ \sqrt{ 16g_1^2 - \left[ (\Gamma-\gamma) + 2i(\Delta_{\rm sg}-\Delta_{\rm ac}) \right]^2 } }{4}, \nonumber\\	  
	\tau_{-}   &=& \frac{ -2(\Delta_{\rm sg}-\Delta_{\rm ac}) + i(\Gamma-\gamma) }{4g_1}
	- \frac{ \sqrt{ 16g_1^2 - \left[ (\Gamma-\gamma) + 2i(\Delta_{\rm sg}-\Delta_{\rm ac}) \right]^2 } }{4},		                         
\end{eqnarray}
and
\begin{eqnarray}
	\mu_1 &=& \frac{1}{\tau_{+}-\tau_{-}}, \quad \mu_2 = \frac{\tau_{+}}{\tau_{+}-\tau_{-}}, \quad
	\mu_3 = \frac{\tau_{-}}{\tau_{+}-\tau_{-}}, \nonumber\\
	\alpha_1 &=& \omega_{+} + \omega_{+}^{*}, \quad \alpha_2 = \omega_{+} + \omega_{-}^{*},  \nonumber\\
	\alpha_3 &=& \omega_{-} + \omega_{+}^{*}, \quad \alpha_4 = \omega_{-} + \omega_{-}^{*}.      
\end{eqnarray}
Here, we assume that the initial state of acoustic phonons is ground state.
In order to quantify the entanglement between two systems, we consider the logarithmic negativity,
which is defined as $E_{\mathcal{N}}=\max[0,-\ln(2\lambda_{-})]$~\cite{plenio2005,vitali2007}. $\lambda_{-}$ is the minimal symplectic
eigenvalue of the covariance matrix $\mathcal{V}$ between two systems under a partial transposition and
can be given by
\begin{eqnarray}
	\lambda_{-} = \sqrt{\frac{ \Sigma(\mathcal{V}) - \sqrt{ \Sigma^2(\mathcal{V}) - 4\det[\mathcal{V}] } }{2}},
\end{eqnarray}
where
\begin{eqnarray}
	\mathcal{V} &=& \left[\begin{array}{cc}
		A   & C \\
		C^T & B
	\end{array} \right], \nonumber\\
	\Sigma(\mathcal{V}) &=& \det[A] + \det[B] - 2\det[C].
\end{eqnarray}
The general criterion of entanglement for bimodal Gaussian states requires the condition
$E_{\mathcal{N}}>0$, which is equivalent to $\lambda_{-}<1/2$. Therefore the logarithmic 
negativity between the idler photons and acoustic phonons during the writing process can 
be calculated as follows
\begin{eqnarray}
	E_{\mathcal{N}} &=& \max\left[ 0, -\ln(2\lambda_{-}) \right], \nonumber \\
\end{eqnarray}
with
\begin{eqnarray}
	\lambda_{-} &=& \sqrt{\frac{ \Sigma\mathcal{V} 
			- \sqrt{ (\Sigma\mathcal{V})^2 - 4\det[\mathcal{V}_{\rm id, b}(t)] } }{2}}, \nonumber\\
	\mathcal{V}_{\rm id, b}(t) &=& \left[ 
	\begin{array}{cc}
		A   &  C \\
		C^T & B
	\end{array} \right], \nonumber\\
	\Sigma\mathcal{V} &=& \det[A] + \det[B] - 2\det[C].
\end{eqnarray}
This minimal symplectic eigenvalue $\lambda_{-}$ can be approximated as follows
\begin{eqnarray}
	\lambda_{-} \approx \frac{1}{2} \times \frac{ 1 + 2n_{\rm th}\left( 1 - e^{-\frac{\Gamma}{2}t} \right)
		+ \frac{ \Gamma n_{\rm th} }{2g_1} e^{-\frac{\Gamma}{2}t} \sin(2g_1t)  }
	{ 1 + e^{-\frac{\Gamma}{2}t} \sin^2(g_1t) }. 
\end{eqnarray}
Since $\lambda_{-}$ experiences a Rabi oscillation with frequency $\sim 2g_1$ in the strong coupling regime, 
its minimum value can be obtained at the optimal time $\tau_1\approx\pi/(2g_1)$ and can be expressed as
\begin{eqnarray}
	\lambda_{-}^{\rm min} \approx \frac{ 1 + \frac{\pi\Gamma n_{\rm th}}{2g_1} }{4}.
\end{eqnarray}
Thus the corresponding maximum value of $E_{\mathcal{N}}$ can be written as follows
\begin{eqnarray}
	E_{\mathcal{N}}^{\rm min} \approx -\ln\left[ \frac{1}{2}\left( 1 + \frac{ \pi\Gamma n_{\rm th} }{2g_1} \right) \right].
\end{eqnarray}
In addition, we use the fidelity to quantify the state transfer between signal photons and acoustic phonons during
the writing process. The fidelity between different two-mode Gaussian states with zero mean can be calculated as follows~\cite{banchi2015}
\begin{eqnarray}
	\mathcal{F}\left[ \mathcal{V}_{\rm re, id}(0), \mathcal{V}_{\rm id, b}(t) \right]
	= \frac{1}{ \sqrt{ \Gamma_{\mathcal{F}} } + \sqrt{ \Lambda_{\mathcal{F}} } 
		- \sqrt{ \left( \sqrt{\Gamma_{\mathcal{F}}} + \sqrt{\Lambda_{\mathcal{F}}} \right)^2 - \Delta_{\mathcal{F}}}  },
\end{eqnarray}
where
\begin{eqnarray}
	\Delta_{\mathcal{F}} &=& \det \left[ \mathcal{V}_{\rm re, id}(0) + \mathcal{V}_{\rm id, b}(t) \right], \nonumber\\
	\Gamma_{\mathcal{F}} &=& 2^4 \det \left[ \left( J \mathcal{V}_{\rm re, id}(0)  \right) 
	\left( J \mathcal{V}_{\rm id, b}(t) \right) - \frac{1}{4}E \right], \nonumber\\
	\Lambda_{\mathcal{F}}&=& 2^4 \det\left[ \mathcal{V}_{\rm re, id}(0) + \frac{i}{2}J \right] 
	\det\left[ \mathcal{V}_{\rm id, b}(t) + \frac{i}{2}J \right],
\end{eqnarray}
and
\begin{eqnarray}
	J = \left[ \begin{array}{cccc}
		0  & 1 & 0  & 0 \\
		-1  & 0 & 0  & 0 \\
		0  & 0 & 0  & 1 \\
		0  & 0 & -1 & 0 \\
	\end{array} \right], \quad
	E = \left[ \begin{array}{cccc}
		1  & 0 & 0 & 0 \\
		0  & 1 & 0 & 0 \\	
		0  & 0 & 1 & 0 \\
		0  & 0 & 0 & 1 \\	
	\end{array} \right].		         	         		        	             	
\end{eqnarray}

We switch off the pump at the optimal time $\tau_1$ in the writing process. In the storage process,
the acoustic phonons are driven by the thermal noise. Thus the dynamics can be given by
\begin{eqnarray}
	\frac{d a_{\rm id}}{dt} &=& \left( -\frac{\gamma_{\rm smf}}{2} + i\Delta_{\rm id} \right) a_{\rm id}
	+ \sqrt{\gamma_{\rm smf}} \xi_{\rm id}, \nonumber \\
	\frac{d b}{dt} &=& \left( -\frac{\Gamma}{2} + i\Delta_{\rm ac} \right)b + \sqrt{\Gamma}\xi_{\rm ac}.             	
\end{eqnarray}
By solving above equations, the covariance matrix $\bar{\mathcal{V}}_{\rm id, b}$ of the idler photons and acoustic phonons
during the storage process can be characterized by
\begin{eqnarray}
	\bar{\mathcal{V}}_{{\rm id, b},11}(t) &=& \frac{1}{2} + N_{10} e^{(\lambda_1+\lambda_1^{*})t}, \nonumber\\
	\bar{\mathcal{V}}_{{\rm id, b},13}(t) &=& {\rm Re}\left[ e^{(\lambda_1+\lambda_2)t}
	\langle a_{\rm id}(0) b(0) \rangle \right], \nonumber\\
	\bar{\mathcal{V}}_{{\rm id, b},14}(t) &=& {\rm Im}\left[ e^{(\lambda_1+\lambda_2)t}
	\langle a_{\rm id}(0) b(0) \rangle \right], \nonumber\\	
	\bar{\mathcal{V}}_{{\rm id, b},33}(t) &=& \frac{1}{2} + e^{(\lambda_2+\lambda_2^{*})t} N_{b0}	  
	+ \frac{\Gamma n_{\rm th}}{\lambda_2+\lambda_2^{*}}    
	\left[ e^{(\lambda_2+\lambda_2^{*})t} - 1 \right],          
\end{eqnarray}
with coefficients
\begin{eqnarray}
	N_{10} &=& 2\eta^2 e^{(\lambda_1+\lambda_1^{*})\tau_1}, \nonumber\\
	\langle a_{\rm id}(0) b(0) \rangle &=& \mu_1 e^{\lambda_1\tau_1}
	\left( e^{\omega_{+}\tau_1} - e^{\omega_{-}\tau_1} \right) (-2\eta^2-i\eta), \nonumber\\
	N_{b0} &=& 2\eta^2 \left| \mu_1 \left( e^{\omega_{+}\tau_1} - e^{\omega_{-}\tau_1} \right) \right|^2 \nonumber\\
	&& + \Gamma n_{\rm th} \left[ \frac{|\mu_3|^2}{\alpha_1} \left( e^{\alpha_1\tau_1} - 1 \right)
	- \frac{\mu_3\mu_2^{*}}{\alpha_2} \left( e^{\alpha_2\tau_1} - 1 \right) 
	- \frac{\mu_2\mu_3^{*}}{\alpha_3} \left( e^{\alpha_3\tau_1} - 1 \right)
	+ \frac{|\mu_2|^2}{\alpha_4} \left( e^{\alpha_4\tau_1} - 1 \right)  \right],
\end{eqnarray}
and
\begin{eqnarray}
	\lambda_1 &=& -\frac{\gamma_{\rm smf}}{2} + i\Delta_{\rm id}, \nonumber\\
	\lambda_2 &=& -\frac{\Gamma}{2} + i\Delta_{\rm ac}.
\end{eqnarray}
Thus the logarithmic negativity can be calculated as follows
\begin{eqnarray}
	\bar{E}_{\mathcal{N}} = \max \left[ 0, -\ln(2\bar{\lambda}_{-}) \right],
\end{eqnarray}
where
\begin{eqnarray}
	\bar{\lambda}_{-} &=& \sqrt{ \frac{ \Sigma\bar{\mathcal{V}} 
			- \sqrt{ \left( \Sigma\bar{\mathcal{V}} \right)^2 - 4\det[	\bar{\mathcal{V}}_{{\rm id, b}}] } }{2} }, \nonumber\\
	\Sigma\bar{\mathcal{V}} &=& \det[\bar{A}] + \det[\bar{B}] - 2\det[\bar{C}], \nonumber\\
	\bar{\mathcal{V}}_{\rm id,b} &=& \left[ 
	\begin{array}{cc}
		\bar{A}   & \bar{C} \\
		\bar{C}^T & \bar{B}
	\end{array} \right].
\end{eqnarray}
The corresponding fidelity can be written as
\begin{eqnarray}
	\bar{\mathcal{F}}\left[ \mathcal{F}_{\rm id}(0), \bar{\mathcal{V}}_{\rm id,b}(t) \right]
	&=& \frac{1}{ \sqrt{\bar{\Gamma}_{\mathcal{F}}} + \sqrt{\bar{\Lambda}_{\mathcal{F}}} 
		- \sqrt{ \left( \sqrt{\bar{\Gamma}_{\mathcal{F}}} + \sqrt{\bar{\Lambda}_{\mathcal{F}}} \right)^2 
			- \bar{\Delta}_{\mathcal{F}} } },   \nonumber\\
\end{eqnarray}
with
\begin{eqnarray}
	\bar{\Delta}_{\mathcal{F}} &=& \det \left[ \mathcal{V}_{\rm id}(0) + \bar{\mathcal{V}}_{\rm id,b}(t) \right], \nonumber\\
	\bar{\Gamma}_{\mathcal{F}} &=& 2^4 \det \left[ \left( J \mathcal{V}_{\rm id}(0) \right)
	\left( J \bar{\mathcal{V}}_{\rm id,b}(t) \right) - \frac{1}{4}E \right], \nonumber\\
	\bar{\Lambda}_{\mathcal{F}} &=& 2^4 \det\left[ \mathcal{V}_{\rm id}(0) + \frac{i}{2} J  \right]
	\det\left[ \bar{\mathcal{V}}_{\rm id,b}(t) + \frac{i}{2} J  \right].
\end{eqnarray}

After a storage period $\tau_{\rm s}$, we apply a second pump to the waveguide and thereby transfer the state from
the acoustic phonons to the retrieval photons. The dynamics of the system in the readout process can be described by
\begin{eqnarray}
	\frac{d a_{\rm id}}{dt} &=& \left( -\frac{\gamma_{\rm smf}}{2} + i\Delta_{\rm id} \right) a_{\rm id}
	+ \sqrt{\gamma_{\rm smf}} \xi_{\rm id}, \nonumber \\
	\frac{d a_{\rm re}}{dt} &=& \left( -\frac{\gamma}{2} + i\Delta_{\rm re} \right) a_{\rm re}
	-ig_2 b + \sqrt{\gamma}\xi_{\rm re}, \nonumber\\
	\frac{d b}{dt} &=& \left( -\frac{\Gamma}{2} + i\Delta_{\rm ac} \right)b - ig_2 a_{\rm re} + \sqrt{\Gamma}\xi_{\rm ac}.  
\end{eqnarray}
Solving above equations, the analytical solution of the covariance matrix $\mathcal{V}_{\rm id,re}$ between
idler and retrieval photons can be characterized by
\begin{eqnarray}
	\mathcal{V}_{{\rm id,re},11}(t) &=& \frac{1}{2} + \tilde{N}_{10} e^{(\lambda_1+\lambda_1^{*})t}, \nonumber\\
	\mathcal{V}_{{\rm id,re},13}(t) &=&	\frac{1}{2} e^{\lambda_1t}\tilde{\mu}_1 \left( e^{\tilde{\omega}_{+}t}
	- e^{\tilde{\omega}_{-}t} \right) \langle a_{\rm id}(0) b(0) \rangle 
	+ \frac{1}{2} e^{\lambda_1^{*}} \tilde{\mu}_1^{*} 
	\left( e^{\tilde{\omega}_{+}^{*}t} - e^{\tilde{\omega}_{-}^{*}t} \right)
	\langle a_{\rm id}^{\dagger}(0) b^{\dagger}(0) \rangle, \nonumber\\
	\mathcal{V}_{{\rm id,re},14}(t) &=&	-i \frac{1}{2} e^{\lambda_1t}\tilde{\mu}_1
	\left( e^{\tilde{\omega}_{+}t} - e^{\tilde{\omega}_{-}t} \right) \langle a_{\rm id}(0) b(0) \rangle  
	+ i \frac{1}{2} e^{\lambda_1^{*}} \tilde{\mu}_1^{*}
	\left( e^{\tilde{\omega}_{+}^{*}t} - e^{\tilde{\omega}_{-}^{*}t} \right)
	\langle a_{\rm id}^{\dagger}(0) b^{\dagger}(0) \rangle, \nonumber\\  
	\mathcal{V}_{{\rm id,re},33}(t) &=&	\frac{1}{2} 
	+ \left| \tilde{\mu}_1 \left( e^{\tilde{\omega}_{+}t} - e^{\tilde{\omega}_{-}t} \right)  \right|^2  
	\tilde{N}_{b}(0) \nonumber\\
	&& + \Gamma n_{\rm th} |\tilde{\mu}_1|^2 \left[ 
	\frac{1}{\tilde{\alpha}_1} \left( e^{\tilde{\alpha}_1} - 1 \right)
	- \frac{1}{\tilde{\alpha}_2} \left( e^{\tilde{\alpha}_2} - 1 \right)
	- \frac{1}{\tilde{\alpha}_3} \left( e^{\tilde{\alpha}_3} - 1 \right) 
	+ \frac{1}{\tilde{\alpha}_4} \left( e^{\tilde{\alpha}_4} - 1 \right) \right],               
\end{eqnarray} 
with coefficients
\begin{eqnarray}
	\tilde{\omega}_{+} &=& -\frac{\Gamma+\gamma}{4} + i\frac{ \Delta_{\rm re} + \Delta_{\rm ac} }{2}
	- i\frac{ \sqrt{ 16g_2^2 - \left[ (\Gamma-\gamma) + 2i(\Delta_{\rm re}-\Delta_{\rm ac}) \right]^2 } }{4}, \nonumber\\
	\tilde{\omega}_{-} &=& -\frac{\Gamma+\gamma}{4} + i\frac{ \Delta_{\rm re} + \Delta_{\rm ac} }{2}
	+ i\frac{ \sqrt{ 16g_2^2 - \left[ (\Gamma-\gamma) + 2i(\Delta_{\rm re}-\Delta_{\rm ac}) \right]^2 } }{4}, \nonumber\\		
	\tilde{\tau}_{+} &=& \frac{ -2(\Delta_{\rm re}-\Delta_{\rm ac}) + i (\Gamma-\gamma) }{4g_2}	 
	+ \frac{ \sqrt{ 16g_2^2 - \left[ (\Gamma-\gamma) + 2i(\Delta_{\rm re}-\Delta_{\rm ac}) \right]^2 } }{4g_2}, \nonumber\\   
	\tilde{\tau}_{-} &=& \frac{ -2(\Delta_{\rm re}-\Delta_{\rm ac}) + i (\Gamma-\gamma) }{4g_2}		
	- \frac{ \sqrt{ 16g_2^2 - \left[ (\Gamma-\gamma) + 2i(\Delta_{\rm re}-\Delta_{\rm ac}) \right]^2 } }{4g_2}, \nonumber\\	                                      
\end{eqnarray}
and
\begin{eqnarray}
	\tilde{\mu}_1 &=& \frac{1}{\tilde{\tau}_{+}-\tilde{\tau}_{-}}, \quad 
	\tilde{\mu}_2 = \frac{\tilde{\tau}_{+}}{ \tilde{\tau}_{+}-\tilde{\tau}_{-} }, \quad
	\tilde{\mu}_3 = \frac{\tilde{\tau}_{-}}{ \tilde{\tau}_{+}-\tilde{\tau}_{-} }, \nonumber\\
	\tilde{\alpha}_1 &=& \tilde{\omega}_{+} + \tilde{\omega}_{+}^{*}, \quad
	\tilde{\alpha}_2 = \tilde{\omega}_{+} + \tilde{\omega}_{-}^{*}, \nonumber\\
	\tilde{\alpha}_3 &=& \tilde{\omega}_{-} + \tilde{\omega}_{+}^{*}, \quad 
	\tilde{\alpha}_4 = \tilde{\omega}_{-} + \tilde{\omega}_{-}^{*},
\end{eqnarray}
and initial conditions of the readout process
\begin{eqnarray}
	\tilde{N}_{10} &=& N_{10} e^{(\lambda_1+\lambda_1^{*})\tau_{\rm s}}, \nonumber\\
	\tilde{N}_{b0} &=& e^{(\lambda_2+\lambda_2^{*})\tau_{\rm s}} N_{b0} 
	+ \frac{1}{\lambda_2+\lambda_2^{*}} \left[ e^{(\lambda_2+\lambda_2^{*})\tau_{\rm s}} - 1 \right], \nonumber\\
	\langle a_{\rm id}(0) b(0) \rangle &=& e^{(\lambda_1+\lambda_2)\tau_{\rm s}} 
	\left[ \mu_1 e^{\lambda_1\tau_1} \left( e^{\omega_{+}\tau_1} - e^{\omega_{-}\tau_1} \right) (-2\eta^2 - i\eta) \right], \nonumber\\
	\langle a_{\rm id}^{\dagger}(0) b^{\dagger}(0) \rangle &=& e^{(\lambda_1^{*}+\lambda_2^{*})\tau_{\rm s}} 	
	\left[ \mu_1^{*} e^{\lambda_1^{*}\tau_1} \left( e^{\omega_{+}^{*}\tau_1} - e^{\omega_{-}^{*}\tau_1} \right) (-2\eta^2 + i\eta) \right].      
\end{eqnarray}
Therefore, the logarithmic negativity $\tilde{E}_{\mathcal{N}}$ between the idler and retrieval photons during
the readout process can be calculated as follows
\begin{eqnarray}
	\tilde{E}_{\mathcal{N}} = \max[0, -\ln(2\tilde{\lambda}_{-})],
\end{eqnarray}
where $\tilde{\lambda}_{-}$ is characterized as
\begin{eqnarray}
	\tilde{\lambda}_{-} &=& \sqrt{\frac{ \Sigma\tilde{\mathcal{V}} 
			- \sqrt{ (\Sigma\tilde{\mathcal{V}})^2 - 4\det[\mathcal{V}_{{\rm id,re}}] } }{2}}, \nonumber\\
	\Sigma\tilde{\mathcal{V}} &=& \det[\tilde{A}] + \det[\tilde{B}] - 2\det[\tilde{C}], \nonumber\\  
	\mathcal{V}_{{\rm id,re}} &=& \left[ 
	\begin{array}{cc}
		\tilde{A}   & \tilde{C} \\
		\tilde{C}^T & \tilde{B}
	\end{array} \right].
\end{eqnarray}
If we consider a short storage time $\tau_{\rm s}\ll1/\Gamma$, $\tilde{\lambda}_{-}$ can approximated as follows
\begin{eqnarray}
	\tilde{\lambda}_{-} = \frac{1}{2} \times \frac{ 1 + 2n_{\rm th}\left( 1 - e^{-\frac{\Gamma}{2}t} \right)
		- \frac{\Gamma n_{\rm th}}{2g_2} e^{-\frac{\Gamma}{2}t} \sin(2g_2t) 
		+ \frac{\pi\Gamma n_{\rm th}}{2g_1} e^{-\frac{\Gamma}{2}t} \sin^2(g_2t) }
	{ 1 + e^{-\frac{\Gamma}{2}\tau_1} e^{-\frac{\Gamma}{2}t} \sin^2(g_2t) }.
\end{eqnarray}
The property of Rabi oscillation indicates that the minimum value of $\tilde{\lambda}_{-}$ can be obtained
at the optimal time $\tau_2\approx\pi/(2g_2)$ and can be reduced to
\begin{eqnarray}
	\tilde{\lambda}_{-}^{\rm min} \approx \frac{1}{4} \left[ 1 + \frac{\pi\Gamma n_{\rm th}}{2g_2} 
	+ \frac{\pi\Gamma n_{\rm th}}{2g_1} e^{-\frac{\pi\Gamma}{4g_2}} \right].
\end{eqnarray}
Thus the maximum value of the logarithmic negativity can be simplified to
\begin{eqnarray}
	\tilde{E}_{\mathcal{N}}^{\rm max} \approx - \ln \left[ \frac{1}{2} 
	\left( 1 + \frac{\pi\Gamma n_{\rm th}}{2g_2} + \frac{\pi\Gamma n_{\rm th}}{2g_1} e^{-\frac{\pi\Gamma}{4g_2}}  \right) \right].
\end{eqnarray} 
%%%%%%%%%%%%%%%%%%%%%%%%%%%%%%%

\end{document}